\documentclass{article}
\usepackage[utf8]{inputenc}
\usepackage[left=0.8in,top=0.8in,right=0.8in,bottom=0.8in]{geometry}
\def\ri{\mathrm i}

\newcommand{\be}{\begin{equation}}
\newcommand{\ee}{\end{equation}}
\newcommand{\bea}{\begin{eqnarray}}
\newcommand{\eea}{\end{eqnarray}}
\newcommand{\ba}{\begin{array}}
\newcommand{\ea}{\end{array}}
\newcommand{\rf}[1] {(\ref{#1})}

\newcommand{\cn}{\mbox{cn}}
\newcommand{\sn}{\mbox{sn}}
\newcommand{\dn}{\mbox{dn}}

\usepackage{Sam}	

\usepackage{enumitem}
\usepackage{authblk}
\usepackage{bm}
\usepackage{amsthm}
\usepackage{amsmath}
\usepackage{accents}
\newcommand{\bdot}[1]{\accentset{\mbox{\large\bfseries .}}{#1}}

\title{Interaction-Phase Dynamics and Spectral Organization in Damped Higher-Order Nonlinear Schr\"odinger Models}
\author{C.M. Schober\thanks{Corresponding author: cschober@ucf.edu} \hspace{6pt} \\
School of Data, Mathematical, and Statistical Sciences, University of Central Florida}

\date{\today}

\begin{document}

\maketitle

\begin{abstract}
  We investigate the dynamical mechanisms underlying the contrasting nonlinear Floquet spectral evolutions observed in viscously damped  and in
  nonlinear mean-flow damped higher-order nonlinear Schr\"odinger models.
  A reduced five-mode carrier-sideband truncation is derived and reformulated
  in amplitude--phase variables to isolate the principal interaction phases associated with the dominant  four-wave interaction products.
  Within this framework, viscous damping acts  modewise and does not
  contribute explicitly to the leading interaction-phase equations,
  whereas nonlinear mean-flow damping acts through the coupled mean-flow
  interactions and, in  the carrier-sideband regime, generates
  terms of the form
  $-\kappa_j \sin(\psi_j)$. The reduced system is used as a mechanism-identification model rather than as a quantitatively accurate
  approximation of the full PDE.
  
To interpret the resulting interaction-phase evolution, we examine recurrent NLS benchmark solutions whose modulation dynamics and Floquet spectrum are independently characterized.
These benchmarks show that substantial interaction-phase evolution
may coexist with recurrent modulation and invariant Floquet-band structure.
Full-PDE diagnostics are then used to compare the phase dynamics within the spectral regimes previously identified for the two damped systems.
Under nonlinear mean-flow damping, the interaction phases develop a persistent large-scale evolution, with the larger phase changes remaining tied to  recurrent focusing events while the Floquet-band configuration persists.
Under viscous damping, the large-scale phase trends change repeatedly without a systematic association with recurrent focusing, while the Floquet spectrum undergoes repeated critical-point crossings and band reconfiguration.
A Fourier-energy assessment further shows that this contrast cannot be attributed simply to broader excitation of Fourier modes in the viscous system. Instead, the results are consistent with a structural distinction in how the two dissipative mechanisms act on the dominant carrier--sideband interactions.

 \end{abstract}

\section{Introduction}

Two complementary perspectives have shaped the modern theory of nonlinear dispersive waves: the dynamical description of resonant mode interactions and the development of soliton and inverse-scattering theory. Early work of David J. Benney established that complex nonlinear wave evolution can often be reduced to the interaction of a finite number of dominant modes, providing a systematic framework for describing nonlinear energy transfer through  low-dimensional dynamical systems \cite{Benney1966,BenneyNewell1967}. In the context of surface gravity waves, reduced evolution equations derived directly from the Euler equations show how nonlinear boundary conditions generate coupled mode interactions and constrained energy exchange among resonant modes \cite{Benney1966,BenneyNewell1967}. These finite-mode interaction systems established
a dynamical framework for modulational instability, sideband growth, nonlinear focusing and recurrent wave evolution \cite{BF1967,Whitham}.

This dynamical viewpoint was subsequently recast into a Hamiltonian framework in which resonant four-wave interactions govern the transfer of energy and specific phase combinations govern the direction and strength of nonlinear exchange \cite{Zakharov,Hasselmann}. In this formulation, nonlinear wave evolution is organized not only by modal amplitudes but also by the relative phases among interacting modes.
Such amplitude–phase descriptions remain central in studies of modulational instability, weak turbulence, and reduced wave-interaction models in both hydrodynamic and optical settings \cite{Janssen,OOSB2001,Trillo-etal2018}.
 In this setting, the interaction phases provide a natural description of coherent nonlinear energy exchange among resonant modes.

A complementary spectral perspective emerged through the development of inverse-scattering theory  for nonlinear evolution equations. The work of Zakharov and  Shabat established the nonlinear Schr\"odinger (NLS) equation as an integrable system whose dynamics can be characterized by invariant nonlinear  scattering data \cite{ZS1972}.
This framework was subsequently generalized by M. J. Ablowitz and collaborators through the Ablowitz–Kaup–Newell–Segur (AKNS) formalism, extending the inverse-scattering methods to broad classes of nonlinear wave equations \cite{AKNS,ablowitz_segur,ablowitz2011} and providing the foundation for later developments in periodic inverse spectral and finite-gap theory, in which coherent nonlinear solutions are characterized through a finite collection of nonlinear spectral bands \cite{AC1991, kodama}.

For spatially periodic problems, the corresponding nonlinear spectral
description  arises through inverse spectral theory and Floquet analysis of the associated Lax  operator. In this setting, coherent nonlinear wave fields are characterized by Floquet spectral data whose band structure encodes the underlying finite-gap dynamics \cite{DubrovinNovikov, McKeanVanMoerbeke}. Subsequent developments in finite-gap and algebraic–geometric integration further established the correspondence between multi-phase nonlinear wave solutions
and finite-band Floquet spectral data \cite{Novikov, ItsMatveev}.

Together, these developments established complementary dynamical and spectral
frameworks for coherent nonlinear wave evolution in integrable systems. On the one hand, finite-mode interaction theories emphasized the role of resonant energy exchange and relative phase dynamics in organizing nonlinear wave motion. On the other hand, inverse-scattering and finite-gap theories characterized coherent wave evolution through invariant nonlinear spectral data and associated Floquet band structure.

For integrable equations, the nonlinear Floquet spectrum is invariant under the evolution and provides a spectral characterization of the solution through its finite-gap structure.
In weakly perturbed or near-integrable settings, however, this spectral invariance is generally lost, and the evolution of the Floquet spectrum provides a natural diagnostic for how perturbations alter coherent nonlinear  dynamics.
Subsequent studies of perturbed near-integrable  systems developed this perspective  through analyses
of  nonlinear Floquet spectral evolution and the associated spectral instabilities, using spectral band deformations and  critical-point crossings to characterize the onset of chaotic dynamics and the breakdown of coherent invariant structures under perturbation \cite{OMB1986, ForestGoedde, McLaughlinSchober, AHS1996, AS2002, EALBBK2020,HW1995}.
More recently, nonlinear spectral analysis has also been applied to soliton-gas dynamics in physical systems governed by perturbed NLS-type equations \cite{FCSR2025, FBD2024}.

In a recent comparative study of rogue wave formation in dissipative higher-order nonlinear Schr\"odinger models,  nonlinear Floquet spectral diagnostics
revealed a striking structural distinction between viscous  damping and
nonlinear mean-flow damping \cite{SI2025}. Under viscous damping, the Floquet spectrum underwent repeated critical-point crossings and band  reconfiguration,
accompanied by less persistent recurrent focusing.
In contrast,  nonlinear mean-flow damping produced sustained
spectral band separation with the complete absence of critical-point crossings.  Long-lived localized one- and two-band spectral configurations also persisted over extended time intervals.
These observations suggested that the two dissipative mechanisms differ not only in how they dissipate energy, but also in how they influence the organization of the underlying nonlinear modulation dynamics.

Although these contrasting Floquet evolutions are well documented, the  dynamical origin of this distinction is  not immediately
apparent from the governing equations. In particular, the results suggest that the two damping mechanisms act differently on the dominant carrier-sideband interactions over the intermediate-time regime considered here. This motivates a reduced interaction-phase description aimed at identifying how the two forms of dissipation enter those interactions.

To investigate this distinction, we derive a reduced five-mode Fourier system
retaining the dominant carrier–sideband interactions underlying the nonlinear evolution and reformulate it in amplitude–phase variables.
The reduction is used to identify how the two dissipative mechanisms enter the principal interaction-phase equations.
Viscous damping acts through modewise decay and has no direct contribution to these phase equations. On the other hand,  nonlinear mean-flow damping acts through the coupled mean-flow interaction terms and, in the carrier–sideband regime, generates contributions of the form $ -\kappa_j \sin(\psi_j)$
introducing direct dissipative feedback into the dominant interaction dynamics.

This distinction suggests that the contrasting Floquet evolutions need not arise from differences in the number of energetically active Fourier modes alone, but may instead reflect differences in how dissipation acts on the nonlinear interactions among those modes. To assess this point, we introduce in Section~5 a carrier-centered Fourier-energy diagnostic that measures the fraction of the full-PDE energy contained in the modes retained by the reduction and in nearby harmonics.

The interaction phases are interpreted as diagnostics of the evolving carrier–sideband interactions rather than as approximately conserved or phase-locked quantities. To establish how their evolution should be interpreted, we first examine recurrent NLS benchmark solutions whose modulation dynamics and Floquet spectral configurations are independently known. These benchmarks show that substantial cumulative interaction-phase evolution can coexist with recurrent modulation and an unchanged Floquet spectral configuration. This provides the reference needed to distinguish substantial phase evolution from genuine reorganization of the underlying nonlinear state.

The same diagnostics are then computed from the full dissipative PDE solutions and interpreted within the Floquet regimes summarized in Section 2. Under nonlinear mean-flow damping, the interaction phases develop a comparatively persistent large-scale evolution whose larger changes remain tied to recurrent focusing while the Floquet-band configuration persists. Under viscous damping, the large-scale phase trends change repeatedly without a systematic association with recurrent focusing, while the Floquet spectrum undergoes repeated critical-point crossings and band reconfiguration.
The Fourier-energy diagnostics further show that the viscous evolution can remain strongly concentrated in the central carrier-sideband modes even while the Floquet spectrum reconfigures.
Thus, the contrasting nonlinear spectral behavior cannot be explained simply by Fourier-mode concentration. Instead, the full-PDE observations are consistent with the structural distinction identified by the reduced interaction-phase
equations, while the Floquet evolution itself is determined independently from the full PDE.

The remainder of the paper is organized as follows. Section 2 introduces the higher-order NLS models with viscous and mean-flow damping and summarizes the  Floquet spectral  diagnostics and prior numerical observations  that motivate the present analysis.
Section 3 derives the reduced five-mode Fourier system and identifies how the two dissipative mechanisms enter the dominant modal interactions.
 Section 4 reformulates the reduced system in amplitude-phase variables and derives the corresponding interaction-phase equations.
 Section 5 computes these phase diagnostics from the full PDE solutions, first using recurrent NLS benchmarks for interpretation and then comparing the nonlinear mean-flow damped and viscously damped evolutions.
 
\section{Governing Equations and Floquet Spectral Motivation}

\subsection{Governing equations}
  In this paper we  investigate how different dissipative mechanisms alter the Floquet spectral organization and nonlinear interaction dynamics of higher-order nonlinear Schr\"odinger (HONLS) systems. We consider two distinct dissipative extensions of the HONLS equation incorporating either weak viscous damping or nonlinear mean-flow damping. The governing equations, associated Floquet spectral diagnostics, and principal prior numerical observations reviewed in this section motivate the reduced interaction-phase analysis developed in the remainder of the paper. The governing equation is given by

\be
\label{DHONLS}
\ri u_t + u_{xx} + 2u|u|^2 + \ri \Gamma u + \epsilon\left[ 2u (1 + \ri\beta)\mathscr{H} \left(|u|^2_x\right) - 8\ri|u|^2 u_x + \frac{\ri}{2} u_{xxx} + 2\Gamma  u_x \right] = 0,
\ee
where $\epsilon >0$ and $0 < \Gamma, \beta <<1$.
The complex envelope, $u(x,t)$, is assumed to be periodic in space with period $L$. Here
 $\mathscr{H}$ denotes the Hilbert transform, defined by
 $\displaystyle \mathscr{H}(f)(x) := \frac{1}{\pi} p.v. \int_{-\infty}^{\infty} \frac{f(\xi)}{\xi - x} d\xi$.

When $\bm \epsilon = \bm \Gamma = \bm \beta = 0$, equation~\rf{DHONLS} reduces to the integrable NLS equation.
Aspects of the associated Floquet spectral theory, which serves as a diagnostic framework for the near-integrable dynamics considered here, are summarized in the following subsection.

For $\bm \epsilon \neq 0$ with $\bm \Gamma = \bm \beta = 0$, equation~\rf{DHONLS}
reduces to the
conservative HONLS, a Hamiltonian variant of the Dysthe equation,  incorporating higher-order nonlinear and dispersive corrections beyond the standard NLS approximation \cite{GT2011,FD2011}. 
 In the present work, the conservative HONLS dynamics serve primarily as a reference evolution against which the effects of the two dissipative perturbations are compared.

 The numerical experiments and reduced-model analysis developed in this paper are focused on the following two dissipative regimes:
 
 \vspace{6pt}
 
\noindent {\bf Viscous HONLS (V-HONLS)}($\bm \Gamma > 0$,  $\bm \beta = 0$):
Dissipation enters through the linear  damping term $i \Gamma u$
together with the  higher-order dissipative correction 
$2\epsilon\Gamma u_x$.
This model arises  from  weakly viscous extensions of the Euler equations 
and provides a physically consistent description of deep water  wave-train dynamics \cite{CG2016}.

\vspace{6pt}
\noindent {\bf Nonlinear mean-flow damped HONLS  (NLD-HONLS)}($\bm\beta > 0$, $\bm \Gamma = 0$): Dissipation enters through the nonlinear mean-flow term,
$2u\ri\beta\mathscr{H}\left(|u|^2_x\right)$, which modifies the  nonlocal interaction structure through amplitude-dependent coupling  between the wave envelope and
induced mean flow. Unlike viscous damping, this mechanism is strongest in regions of strong modulation where the wave envelope is both large and rapidly varying. Although introduced phenomenologically in Dysthe-type models \cite{UK1994,KO1995,IS2011}, this mechanism has proven effective in modeling localized rogue-wave dynamics in dissipative wave trains.

Both damped models capture frequency downshifting, an important phenomenon observed in laboratory experiments and field measurements. However, previous numerical studies revealed a striking difference in the associated Floquet spectral dynamics: viscous damping produces repeated spectral reconnection and continual critical-point crossings, whereas nonlinear mean-flow damping produces persistently organized Floquet  band evolution without any critical-point crossings \cite{SI2025}.

Understanding the dynamical mechanism underlying these contrasting spectral evolutions is a principal objective of the present work.
The reduced Fourier and interaction-phase analysis developed in the following sections is aimed at identifying how the different dissipation mechanisms modify the dominant 
four-wave interactions underlying the contrasting  Floquet spectral evolutions.

 \subsection{Floquet Spectral Diagnostics}

To characterize the nonlinear spectral structure underlying the dissipative HONLS dynamics, we employ the Floquet spectrum associated with the integrable NLS equation as a diagnostic tool. In the integrable setting, the Floquet spectrum provides the finite-gap representation of the solution and detailed information about the associated nonlinear evolution. Here, ``finite-gap'' refers to solutions whose nonlinear Floquet spectrum is characterized by a finite collection of nontrivial spectral bands, which encode the principal nonlinear degrees of freedom of the solution. For weakly perturbed or near-integrable systems, although the spectrum is no longer invariant, its evolution continues to provide a sensitive diagnostic of the evolving nonlinear interactions.

 The Floquet spectrum is defined through the spatial Zakharov--Shabat (Z-S) spectral problem associated with the Lax pair formulation of the NLS equation \cite{ZS1972},
 \be
   L^{(x)}\mathbf{v}=\begin{pmatrix}\partial_x+i\lambda & -u \\\bar{u} & \partial_x-i\lambda\end{pmatrix}\mathbf{v}=0,\label{ZSsystem}
 \ee
 where \(\lambda\) is the spectral parameter and \(u(x,t)\) serves as the potential in the Z-S problem.
 Let \(\Psi(x;\lambda)\) denote a fundamental solution matrix of the Z-S system.
Since the spectral problem depends on the evolving potential $u(x,t)$, the associated Floquet discriminant is time dependent and is defined by
 \be
   \Delta(u,\lambda)=\mathrm{Trace}\left(\Psi(x+L;\lambda)\Psi^{-1}(x;\lambda)\right),\label{FloquetDisc}
 \ee
 which characterizes the behavior of the eigenfunctions over one spatial period \(L\).

 The Floquet spectrum is defined  in terms of the discriminant by
 \be\sigma(u)=\left\{\lambda\in\mathbb{C}\mid\Delta(u,\lambda)\in\mathbb{R},\ -2\le \Delta(u,\lambda)\le 2\right\}.\label{FloquetSpectrum}
 \ee
 For integrable NLS evolution, both the discriminant \(\Delta(\lambda)\) and the associated Floquet spectrum \(\sigma(u)\) are invariant in time and provide the infinite hierarchy of NLS conservation laws.

 The spectrum for an NLS solution consists of the entire real axis
 together with curves in the complex \(\lambda\)-plane, referred to as spectral bands.
 Special values of \(\lambda\) for which \(\Delta=\pm2\) are referred to as periodic and antiperiodic points, respectively, and will
collectively be referred to here as  periodic points.
 Among these, the simple periodic points form the set
 \be\sigma_s(u)=\left\{\lambda_j^s\mid\Delta(\lambda_j^s)=\pm2,\ \partial_\lambda\Delta\neq0\right\}.
 \label{SimplePeriodicPoints}\ee
 The nonreal simple  periodic points occur in complex conjugate pairs
 and determine the endpoints of the nonreal spectral bands.

 Of particular importance in the present work are degeneracies   of the Floquet spectrum occurring at points \(\lambda_j^c\) satisfying 
 \be
   \frac{\partial \Delta}{\partial \lambda}\Big|_{\lambda_j^c}=0.
   \label{CriticalPoints}
 \ee
 Among these,  double points  \(\lambda_j^d\) are degenerate periodic or antiperiodic points satisfying

 \be
   \Delta(\lambda_j^d)=\pm2,\qquad\frac{\partial \Delta}{\partial \lambda}=0,\qquad\frac{\partial^2\Delta}{\partial \lambda^2}\neq0.\label{DoublePoints}
 \ee
 In the classical near-integrable NLS literature, double-point crossings play a central role in the onset of homoclinic instabilities and chaotic spectral dynamics \cite{EFM1990,McLaughlinSchober,AHS1996}.

 In the higher-order HONLS systems considered here, however, spectral reconfiguration occurs more generally through nonperiodic critical points, for which 
\be
 \frac{\partial \Delta}{\partial \lambda}=0,\qquad\Delta\neq\pm2.
 \ee
 Accordingly, throughout the remainder of this paper, the term ``critical points'' will refer specifically to these nonperiodic critical points, 
 while ``double points'' will refer to degenerate periodic points satisfying \(\Delta=\pm2\).

 \vspace{6pt}
 \noindent {\bf Terminology and Quantitative Criteria.}
 
 In the dissipative HONLS dynamics considered here, spectral bands may meet at the critical points defined above. A critical-point crossing refers to such an event
 when it produces a change in how the bands connect; this change will be referred to as a {\bf band reconnection} or {\bf Floquet spectral reconfiguration}.
 Repeated occurrences of such events will be referred to as repeated or persistent band reconnection.

 Throughout this paper, {\bf Floquet spectral organization}  means that the evolving band configuration remains free of critical-point crossings. Smooth deformation, drift, or lengthening of the bands does not by itself constitute spectral reorganization; a change of spectral state occurs when a critical-point crossing changes how the bands connect.

 To quantify {\bf Floquet spectral localization}, we monitor the lengths of the dominant spectral bands. Following the criterion introduced in \cite{SI2022}, and subsequently used in the broader Floquet study \cite{SI2025},  a band
 \[\gamma(t;\lambda_m,\lambda_n)\] is regarded as Floquet spectrally localized when
 \be|\gamma(t;\lambda_m,\lambda_n)|=|\lambda_m(t)-\lambda_n(t)|<0.025.\label{BandLocalization}
 \ee

 Spectral configurations for which one or two dominant spectral bands satisfy the localization criterion will be referred to here as localized one-band and localized two-band configurations, respectively.
Floquet spectral localization and Floquet spectral organization are distinct: a band may cease to satisfy the localization threshold while the overall band configuration remains organized as long as no critical-point crossing occurs.

  \vspace{6pt}
  \noindent {\bf Numerical Framework  for Near-Integrable Equations.}

  The following paragraphs summarize the numerical framework used to compute and interpret the V-HONLS and NLD-HONLS dynamics examined in Section 2.3. Additional implementation details are given in Section 5.

  Equation \rf{DHONLS} is solved numerically using a Fourier pseudo-spectral discretization in space combined with a fourth-order exponential Runge--Kutta method in time \cite{CM2002}. The Floquet spectra are computed numerically using the method developed in \cite{OMB1986}. After solving the Z-S system, the Floquet discriminant is constructed numerically and the zeros of $\Delta \pm 2$ are determined using a root solver based on Muller's method, allowing numerical reconstruction of the spectral bands. Numerical resolution studies and comparison with the integrable NLS limit confirm that the  computations accurately resolve the band evolution and critical-point dynamics throughout the simulations considered here. 

The numerically observed short-time spectral evolution of the SPB initial data under both the V-HONLS and NLD-HONLS dynamics is consistent with perturbation analysis predicting asymmetric splitting of the complex double points at  ${\mathcal O}(\epsilon)$ \cite{SI2021}.

\vspace{6pt}
\noindent {\sc Initial Data:} The simulations presented below  are
initialized  using two-mode spatially periodic breathers (SPBs) of the integrable NLS equation:
\be
\label{SPB_ic}
  u(x,0) = U(x,t_0,\rho,\tau),
\ee
where $U(x,t,\rho,\tau)$ is given by \rf{comboSPB} in the  Appendix. These initial data represent highly
structured, localized, and steep wave fields and are therefore well suited for examining the subsequent focusing and Floquet spectral evolution. 

\vspace{6pt}
\noindent {\bf Focusing and Rogue-Wave Criterion.}
Following \cite{SI2025}, the strength of a focusing event is measured by
\begin{equation}
  S(t) = \frac{U_{max}(t)}{H_s(t)}, \quad U_{max}(t) = \mbox{max}_{x\in [0,L]} |u(x,t)|,
  \label{rwcond}
\end{equation}
where  $H_s(t)$ is the significant
wave amplitude.  A focusing  event at time $t^*$ is classified as a rogue wave when
\begin{equation}
  S(t^*) \ge 2.2.
  \label{rwcondb}
  \end{equation}
We use recurrent focusing to refer to repeated focusing--defocusing episodes of the wave envelope. In the Floquet study \cite{SI2025}, rogue-wave events satisfying this strength criterion and exhibiting localized one- or two-band spectral configurations according to \rf{BandLocalization} were characterized as one- or two-soliton-like rogue waves, respectively. In the present work, this terminology is used to identify the strongly localized focusing events whose Fourier-energy and interaction-phase dynamics are examined in Section~5.

\vspace{6pt}
\noindent {\bf Floquet  Spectral-Plot Notation.}
In the Floquet spectral plots, periodic and antiperiodic points  are marked by large 
symbols, while the spectral bands are traced by smaller symbols.
A  small ``$\times$'' denotes  $\Delta <0$, with a 
large ``$\bigtimes$'' indicating  $\Delta = -2$.
A small box denotes $\Delta >0$, with a large box indicating $\Delta = 2$.
As an additional visual aid, the sign of $\Delta$ is also color coded:
``$\times$'' symbols are shown in red and boxes in blue.
Thus, symbol size distinguishes periodic/antiperiodic points from other points along the spectral bands, while symbol type and color distinguish the sign of the Floquet discriminant.
Because the NLS Floquet spectrum is symmetric with respect to complex conjugation, only the upper half of the $\lambda$-plane  is shown.

 \subsection{Prior Floquet Spectral Observations}
 The Floquet spectral classifications summarized in this subsection were established in the numerical study \cite{SI2025} and are not rederived here. We recall only the features needed to motivate and interpret the interaction-phase analysis developed in the remainder of the paper. Although the V-HONLS and NLD-HONLS systems evolve from the same initial spectral configuration and exhibit similar short-time dynamics, their subsequent Floquet evolution differs markedly. Under nonlinear mean-flow damping, recurrent rogue-wave focusing is accompanied by long-lived localized one- and two-band configurations and an absence of critical-point crossings. Under viscous damping, the initially localized spectral configurations are transient and the spectrum undergoes repeated critical-point crossings and band reconnections. Thus, the principal distinction used below is between a persistently organized NLD-HONLS Floquet evolution and a repeatedly reconfiguring V-HONLS evolution.
  
\vspace{6pt}
\noindent {\bf  Early-time Floquet spectra for HONLS, V-HONLS, and NLD-HONLS:}
Figure~\rf{early-spec}(a) shows the Floquet spectrum of the initial two-mode SPB data,  which  coincides with that of the underlying Stokes wave and consists of a single spectral band together with two imaginary double points.
Spectral symbols and colors throughout Figures~\rf{early-spec}, \rf{VHONLS_spec}, and \rf{NLD_SPB_spec} follow the Floquet spectral-plot notation defined in Section~2.2.
In all three models, the short-time spectral evolution is nearly identical through the first rogue-wave focusing event near $t \approx 3.5$,
which satisfies the rogue-wave strength criterion \rf{rwcondb}.
During this interval, the two imaginary double points  split
asymmetrically from the imaginary axis, as seen in Figure~\rf{early-spec}(b), consistent with the perturbation analysis of \cite{SI2021}.
The resulting nonreal  spectral bands then contract, producing a localized
one-band configuration and subsequently  a localized two-band configuration
shown in   Figure~\rf{early-spec}(c)--(d), according to the Floquet
spectral-localization criterion \rf{BandLocalization}.
Thus, no distinction between the two damping mechanisms is apparent during this initial interval: the focusing and Floquet-band evolution are essentially the
same for  the HONLS, V-HONLS, and NLD-HONLS systems. Their contrasting spectral behavior emerges during the subsequent evolution. The same SPB initial data are used in Section~5, where the present study augments these previously established Floquet and focusing classifications with Fourier-energy and interaction-phase diagnostics computed from the full PDE solutions.

  \begin{figure}[hpt!]
  \centerline{
    \includegraphics[width=0.33\textwidth]{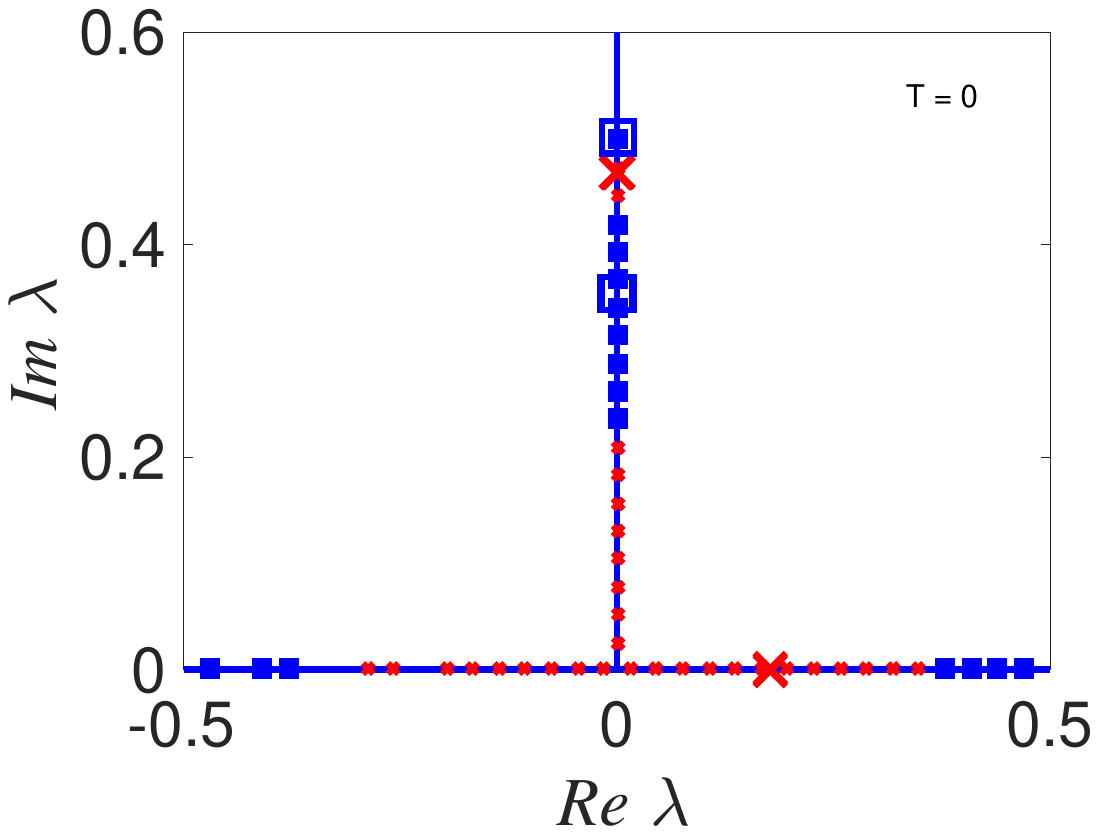}
    \hspace{6pt}\includegraphics[width=0.33\textwidth]{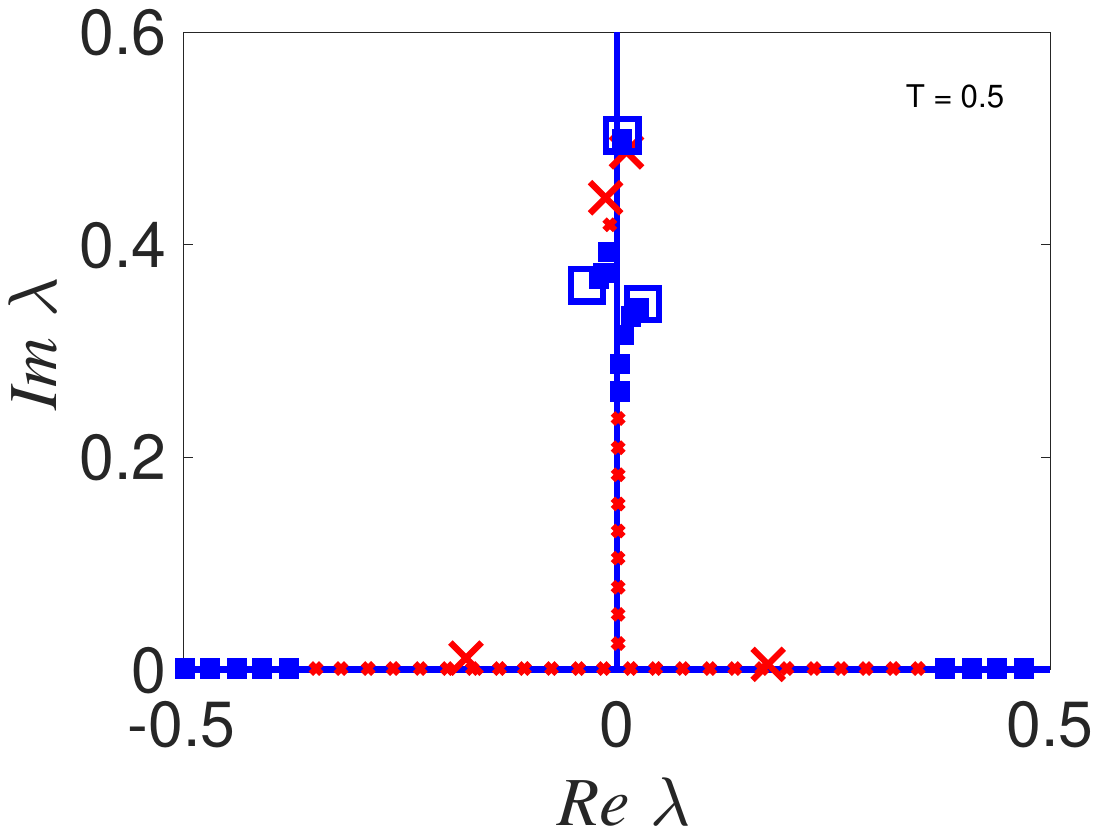}
  }
  \vspace{-48pt}
  \centerline{(a)\hspace{0.33\textwidth}(b)}
    \centerline{
\includegraphics[width=0.33\textwidth]{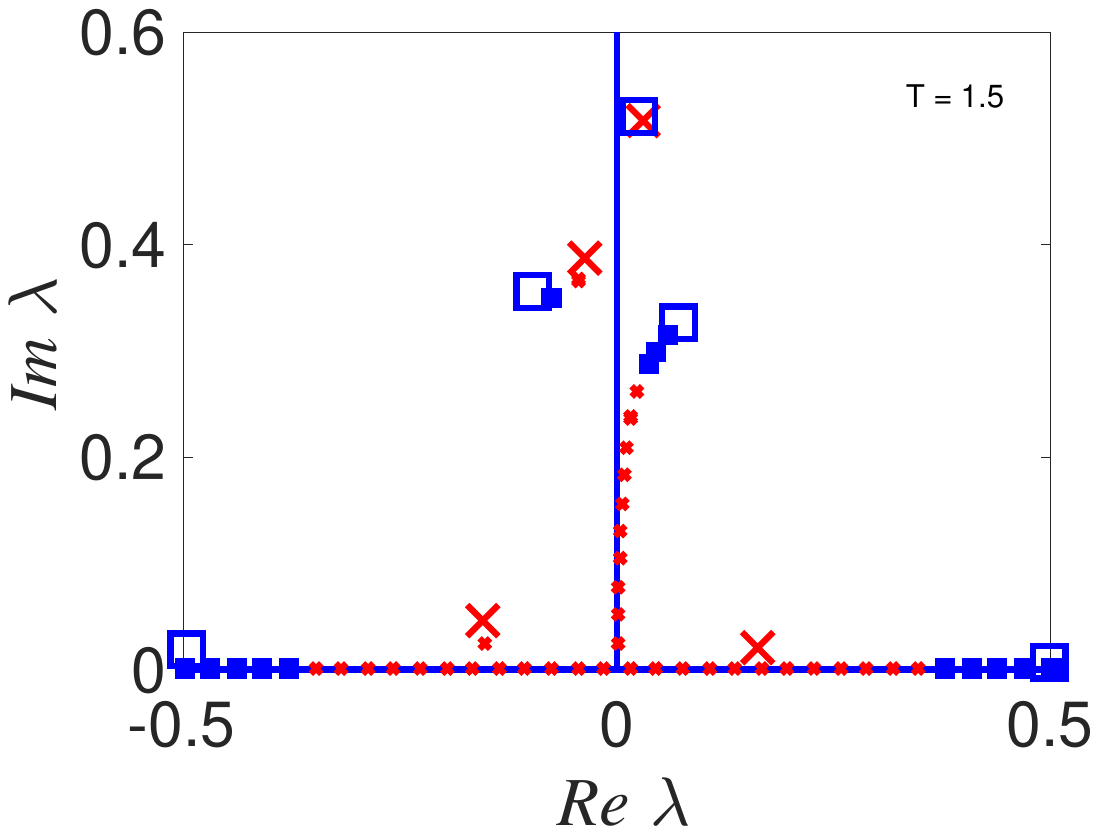}
\hspace{6pt}\includegraphics[width=0.33\textwidth]{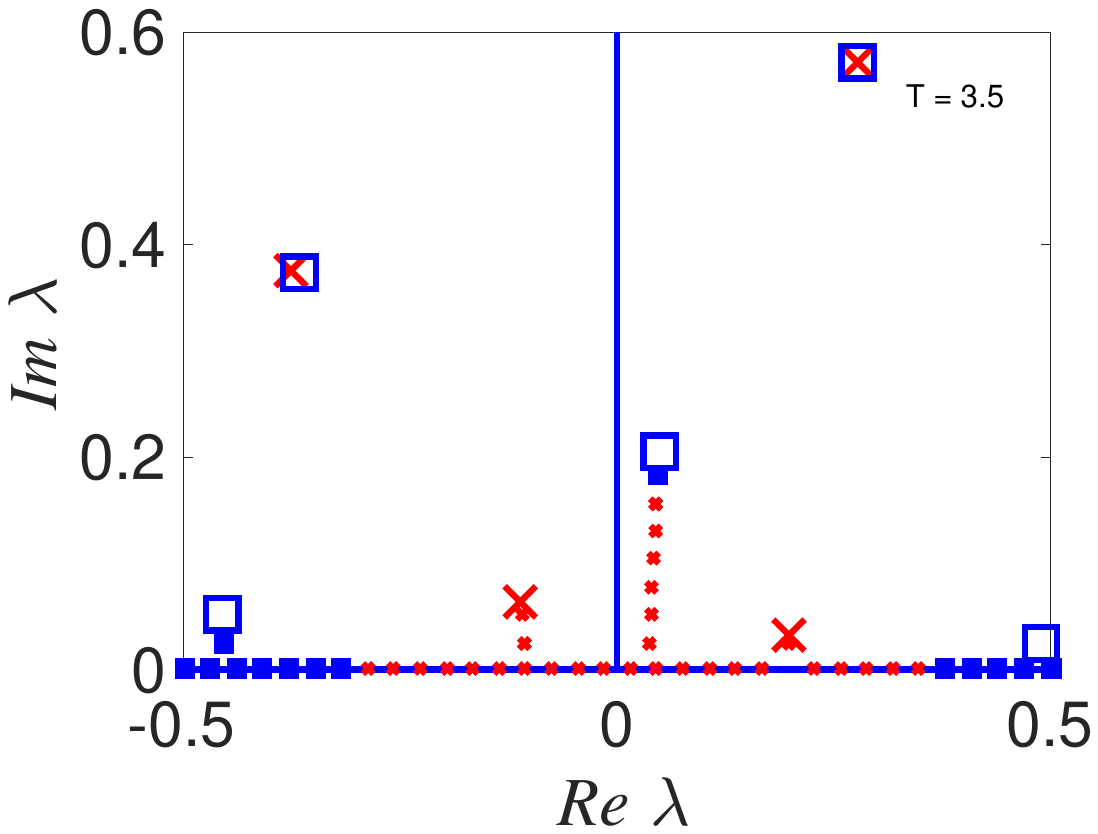}
  }
  \vspace{-48pt}
  \centerline{(c)\hspace{0.33\textwidth}(d)}
  \caption{Early-time Floquet spectra for the SPB initial data under HONLS, V-HONLS, and NLD-HONLS at (a) $t = 0$, (b) $t=0.5$, (c) $t=1.5$, and (d) $t=3.5$.  The initial imaginary double points split asymmetrically and the resulting bands contract into the localized two-band configuration in (d), as defined by the Floquet spectral-localization criterion \rf{BandLocalization}.
    The time $t \approx 3.5$ corresponds to the first rogue-wave focusing event as defined by the strength criterion \rf{rwcondb}. Spectral symbols and colors follow the Floquet spectral-plot notation of Section~2.2. The three models exhibit essentially the same Floquet evolution over this initial focusing interval.}
  \label{early-spec}
  \end{figure}

\vspace{6pt}
\noindent {\bf V-HONLS spectral evolution:}
In the viscous model, the initially localized Floquet configurations are only transient. Following the first rogue-wave focusing event near
$t \approx 3.5$, the spectral bands begin to expand and subsequently undergo repeated critical-point crossings and band reconnections.
Figure~\rf{VHONLS_spec}(a)-(c) illustrates one representative crossing event near
$t \approx 54.7$,
where the change in how the bands connect constitutes a Floquet spectral reconfiguration in the terminology of Section~2.2. Although localized one-band configurations reappear intermittently during the evolution, they persist only for limited time intervals before further critical-point crossings occur. The V-HONLS evolution is therefore characterized by repeated changes in the Floquet-band configuration rather than by persistent Floquet spectral organization.
  \begin{figure}[hpt!]
  \centerline{
\hspace{6pt}\includegraphics[width=0.33\textwidth]{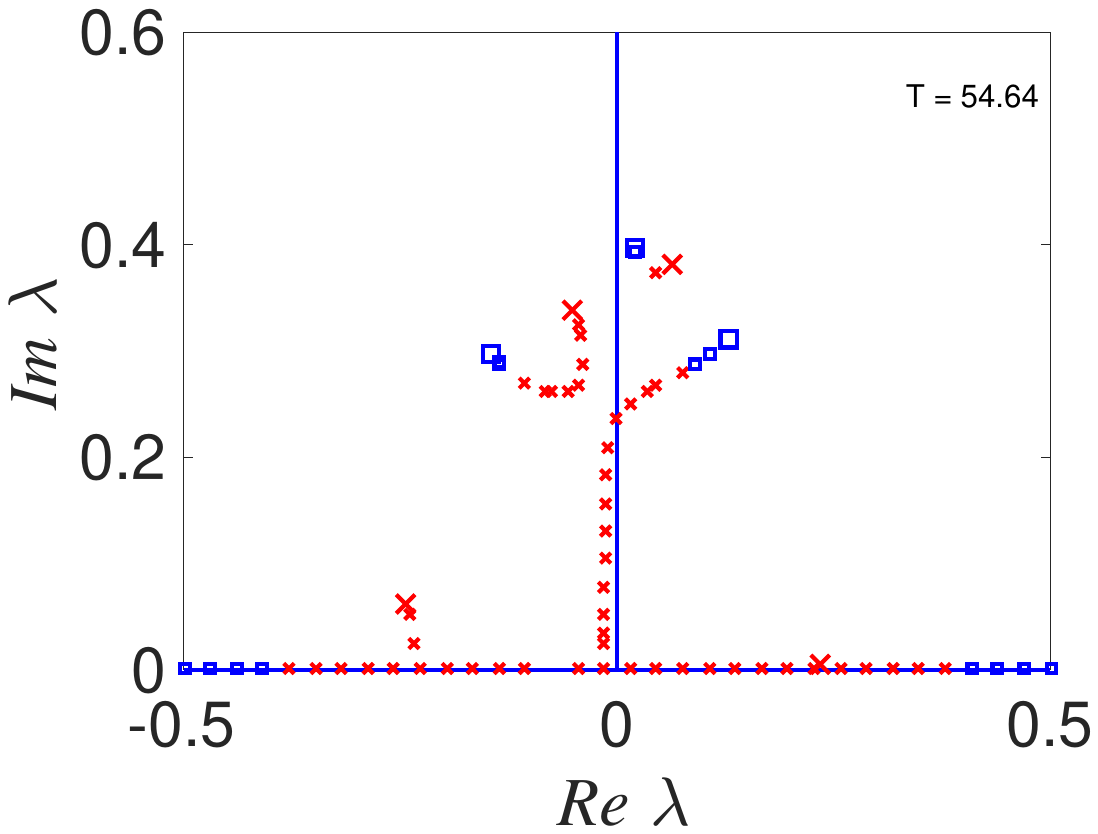}
\hspace{6pt}\includegraphics[width=0.33\textwidth]{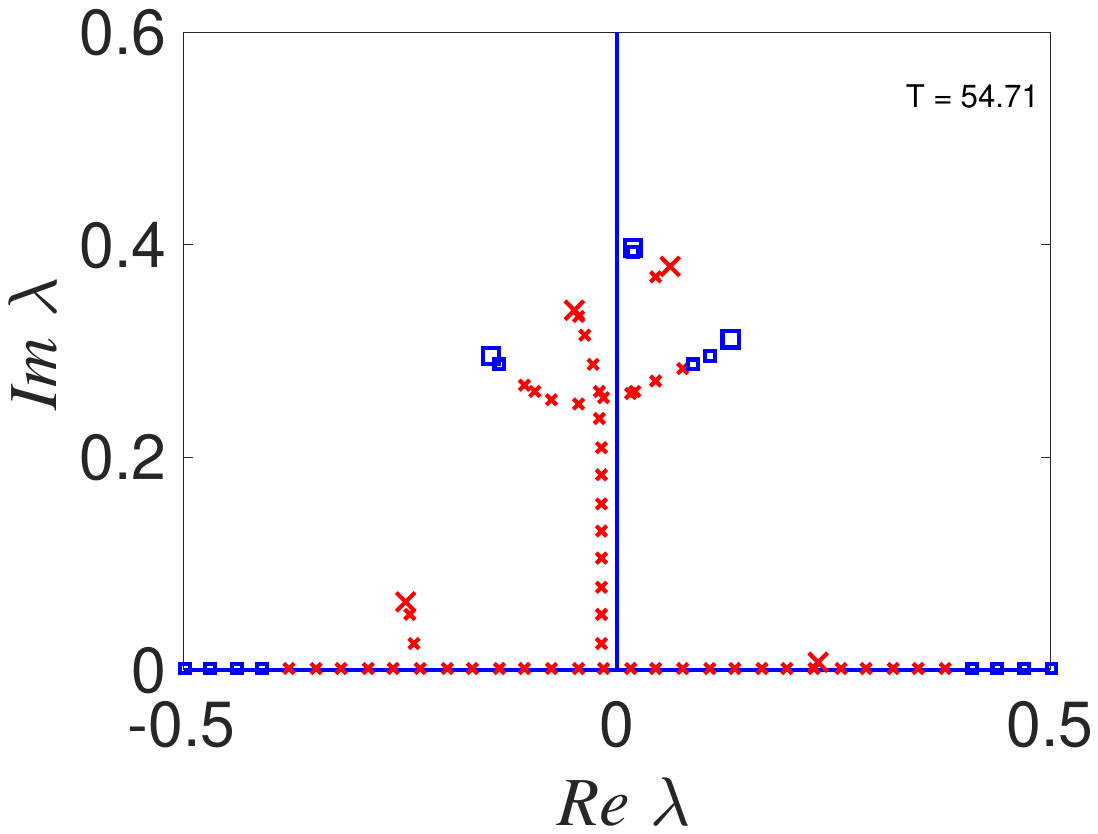}
\hspace{6pt}\includegraphics[width=0.33\textwidth]{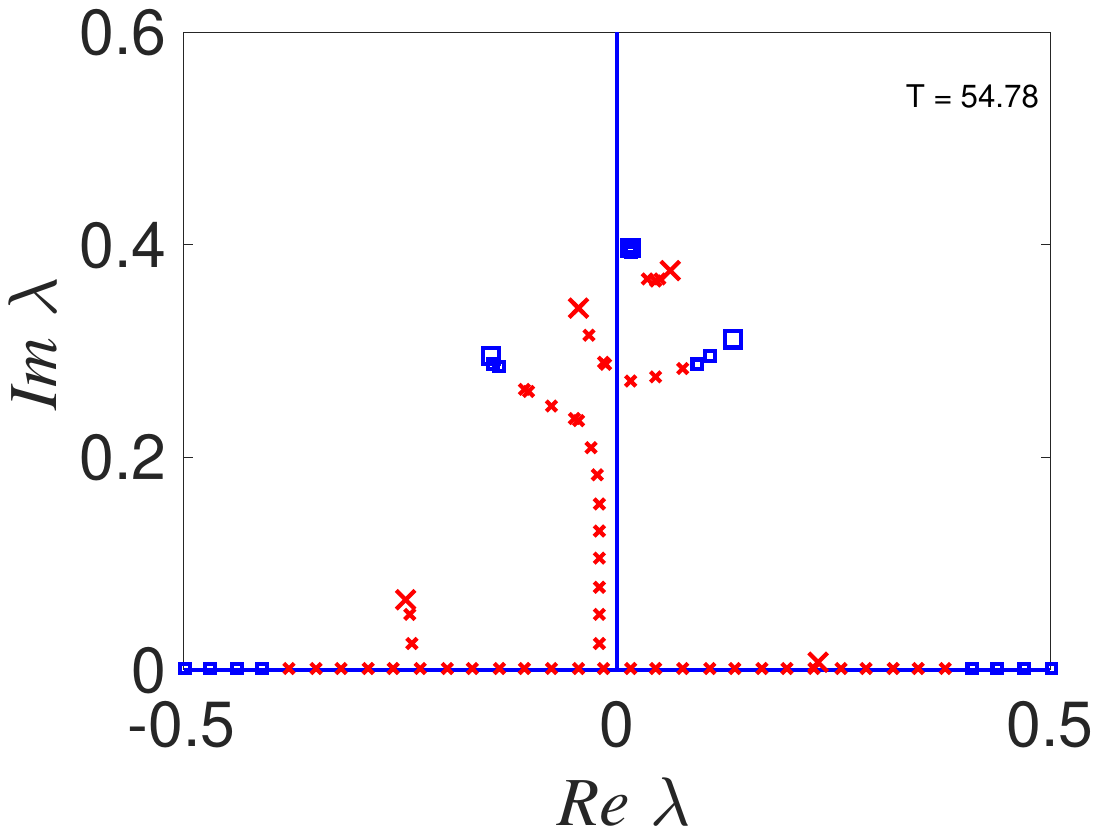}
  }
    \vspace{-48pt}
  \centerline{(a)\hspace{0.33\textwidth}(b)\hspace{0.33\textwidth}(c)}
  \caption{V-HONLS Floquet spectra for $\Gamma = .002$ at
    (a) $t = 54.64$, (b) $t =  54.71$, and (c) $t = 54.78$
    illustrating a representative critical-point crossing at
    $t \approx 54.7$ and the associated change in how the spectral bands connect. This band reconnection constitutes Floquet spectral reconfiguration as defined in Section~2.2. Spectral symbols and colors follow the Floquet spectral-plot notation of Section~2.2.
    }
  \label{VHONLS_spec}
  \end{figure}

  To place the representative crossing in Figure~\rf{VHONLS_spec}(b)  within the broader parameter study reported in \cite{SI2025},
  Figure~\rf{crossings} summarizes the times at which critical-point crossings occur as the viscous damping parameter  $\Gamma$ is varied. The horizontal axis gives $\Gamma$, with $\Gamma = 0$   corresponding to the conservative HONLS limit, while the vertical axis gives the time of each detected crossing. Each marker therefore represents a
  critical-point crossing in a particular evolution rather than a point on a Floquet spectral band.
  Blue dots denote crossings involving real critical points, while red
  $\times$  symbols denote crossings involving complex critical points. The figure thus provides a parameter-time plot showing that the crossing illustrated in Figure~\rf{VHONLS_spec}(b) is part of a broader pattern of repeated Floquet spectral reconfiguration across the HONLS and V-HONLS evolutions.
  \begin{figure}[hpt!]
  \centerline{
\includegraphics[width=0.4\textwidth]{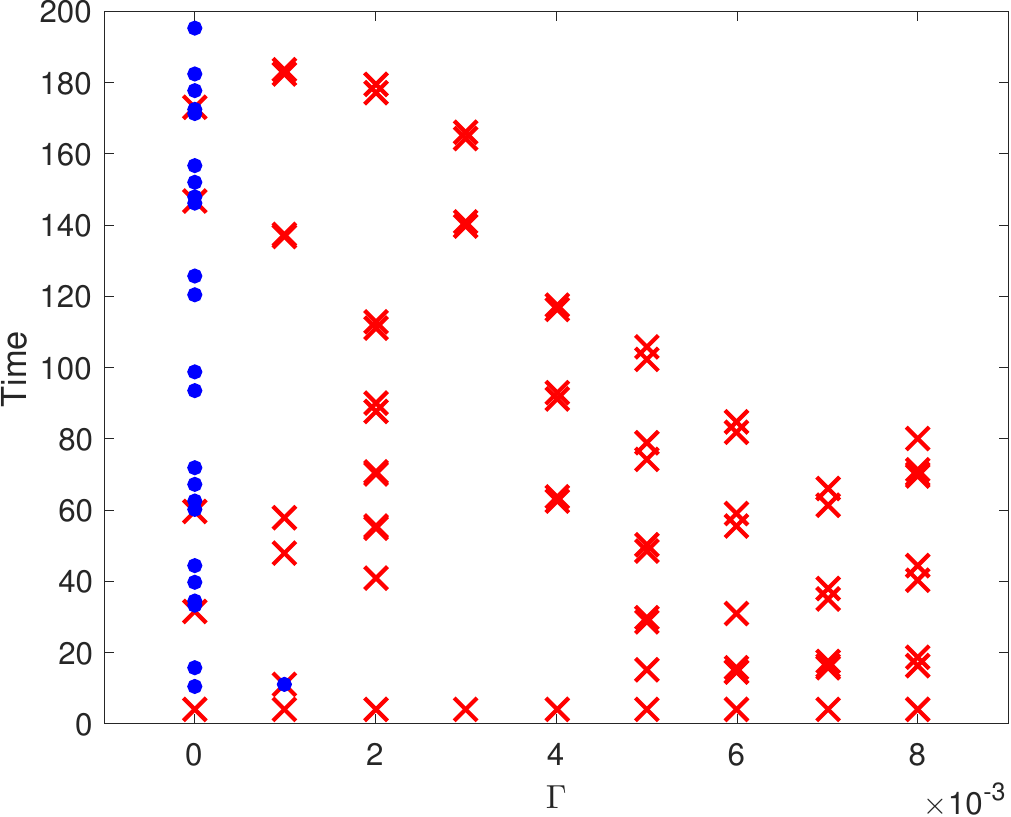}
  }
  \caption{Parameter-time plot of critical-point crossings for the HONLS
    ($\Gamma = 0$) and V-HONLS ($\Gamma > 0$) evolutions with common SPB initial data. The horizontal axis gives the viscous damping parameter $\Gamma$,    and the vertical axis gives the crossing time. Blue dots denote
    crossings involving real critical points, while red $\times$ symbols denote crossings involving complex critical points. }
  \label{crossings}
  \end{figure}
 
\vspace{6pt}
  \noindent {\bf  NLD-HONLS spectral evolution:}
  In contrast to the HONLS and V-HONLS dynamics, nonlinear mean-flow damping produces a persistently organized  long-time Floquet spectral evolution.
  Following the first rogue wave focusing  event near  $t \approx 3.5$, the localized two-band spectral configuration formed during the common early-time evolution persists for an extended time interval. 
  Figure~\rf{NLD_SPB_spec}(a)-(c) shows representative spectra at $t = 15, 23$, and $47$.  During this evolution, the dominant spectral bands slowly expand in length
  and drift as the wave amplitude decays, while the Floquet spectrum remains highly organized  and free of critical-point crossings.
  By 
  $t \approx 47$, Figure~\rf{NLD_SPB_spec}(c), one band has exceeded the Floquet spectral-localization threshold \rf{BandLocalization}, while the other remains localized.
  This loss of localization does not constitute a change in Floquet spectral
  configuration.

   The broader parameter study of \cite{SI2025} showed that the absence of 
   critical-point crossings is a robust feature of the NLD-HONLS dynamics
   rather than a  feature specific to  the $\beta = 0.1$ evolution shown in
   Figure~\rf{NLD_SPB_spec}. Across the range  $0<\beta\leq 0.8$ examined in
   the study, no critical-point crossings were detected. 

   Thus, although the dominant bands may gradually expand and one may cease to satisfy the Floquet spectral-localization criterion, the NLD-HONLS evolution remains within a persistently organized Floquet regime over the parameter range examined. This behavior contrasts sharply with the repeated critical-point crossings and spectral reconfigurations observed in the HONLS and V-HONLS evolutions.
 
      \vspace{-48pt}
\begin{figure}[htp!]
\centerline{
    \includegraphics[width=0.33\textwidth]{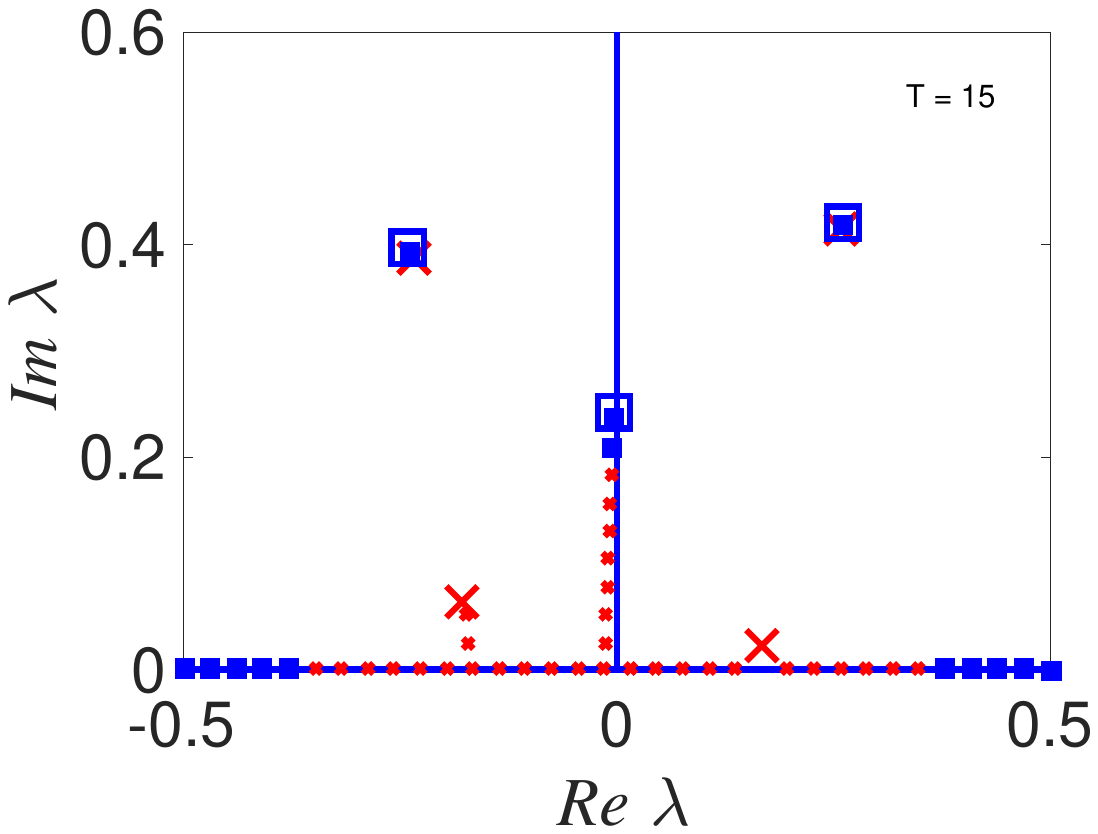}
\hspace{6pt}\includegraphics[width=0.33\textwidth]{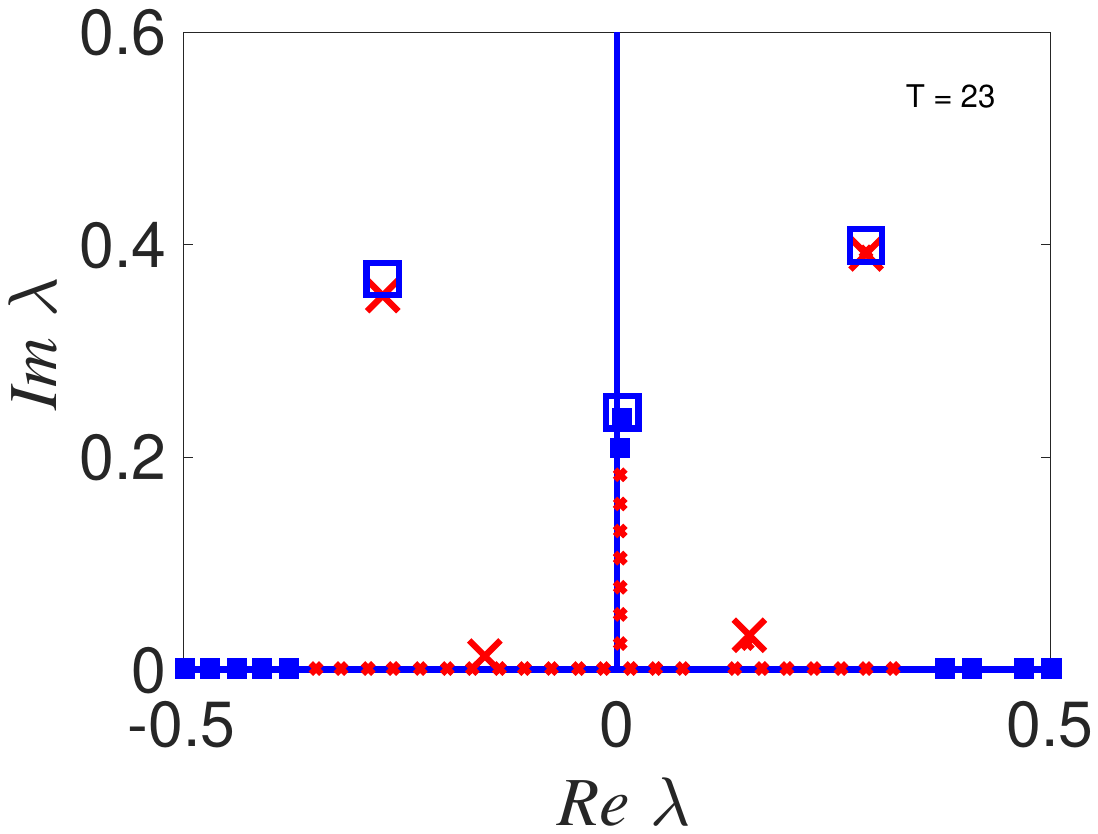}
\hspace{6pt}\includegraphics[width=0.33\textwidth]{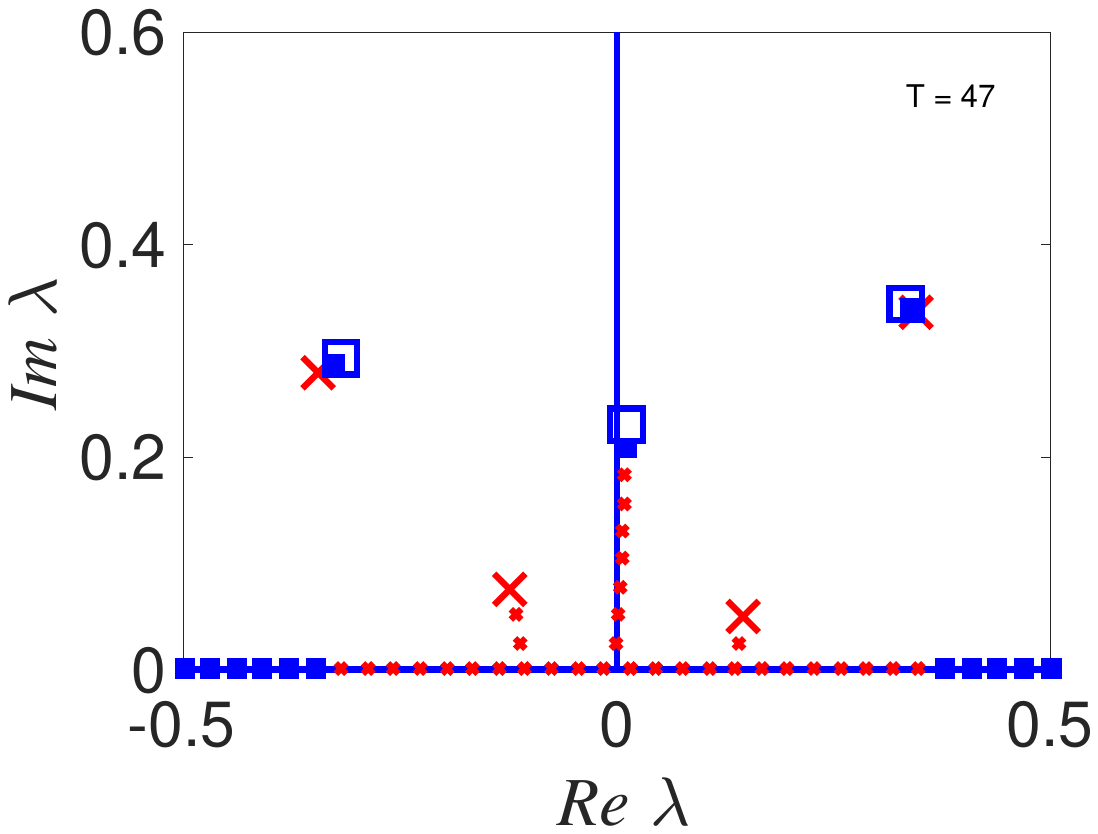}
}  
    \vspace{-48pt}
  \centerline{(a)\hspace{0.33\textwidth}(b)\hspace{0.33\textwidth}(c)}
\caption{NLD-HONLS Floquet spectra for $\beta = 0.1$
  at  (a) $t = 15$, (b) $t= 23$, and
  (c) $t=47$, illustrating the 
  persistence of the Floquet-band configuration during the dissipative
  evolution.  The dominant bands slowly expand and drift without critical-point crossings. By $t = 47$, one band has exceeded the Floquet spectral-localization threshold \rf{BandLocalization}, while the other remains localized. This loss of localization does not constitute a change in Floquet spectral configuration. Spectral symbols and colors follow the Floquet spectral-plot notation of Section~2.2.
}
\label{NLD_SPB_spec}
\end{figure}

A central conclusion of the previous numerical study \cite{SI2025} was therefore structural rather than merely phenomenological: the two damping mechanisms differ not only in how they dissipate energy, but also in how they affect the long-time organization of the nonlinear Floquet spectrum.
Under nonlinear mean-flow damping, the spectral configuration remains free of critical-point crossings even as the dominant bands expand and drift, whereas the HONLS and V-HONLS evolutions undergo repeated changes in spectral configuration.

The dynamical mechanism underlying this contrast is not apparent from the Floquet spectra alone. To investigate how the two forms of dissipation act on the dominant nonlinear interactions, we next derive a reduced five-mode Fourier system capturing the principal carrier-sideband interactions and analyze the associated interaction-phase dynamics.

\section{Five-Mode Formulation}

To examine the dominant nonlinear interactions underlying the dissipative
evolution,
 we introduce a reduced Fourier description that retains the minimal mode-coupling structure necessary to capture  the leading four-wave mixing processes and their associated interaction-phase dynamics. The reduction is used primarily to identify how the two damping mechanisms enter these interactions, rather than as a quantitatively complete approximation of the full PDE or its nonlinear Floquet spectrum.

Modulational energy exchange is organized primarily through the carrier wave and its nearest sidebands, while nonlinear coupling among the first sidebands generates second-order harmonics. This leads naturally to a five-mode Fourier truncation consisting of the carrier together with the first two pairs of sidebands.

Specifically we consider  the five-mode Fourier truncation

\be
u(x,t)=\sum_{n=-2}^{2} A_n(t)e^{in\mu x},
\qquad
\mu=\frac{2\pi}{L},
\label{five}
\ee
where $L$ is the spatial period. The mode $n=0$ corresponds to the carrier mode, while $n = \pm 1$  and $n = \pm 2$ denote the first and second symmetric sideband pairs.
The inclusion of the $\pm 2$  sidebands is required by the nonlinear coupling structure of the governing equation, since interactions among the first sidebands generate second-order harmonics. The resulting system is therefore the smallest closed Fourier truncation capable of capturing the leading cubic interactions, higher-order corrections, and nonlocal mean-flow coupling associated with the dominant modulation dynamics.

A symmetry constraint is not imposed on the amplitudes $A_n(t)$  since the higher-order and dissipative terms generally break the reflection symmetries present in idealized NLS reductions. In particular, dissipative downshifting produces asymmetric sideband evolution that cannot be captured within a symmetry-restricted truncation. We therefore allow the sideband amplitudes and phases to evolve independently and adopt the following: 
\[
A_j(t)=0
\qquad \text{for }  |j| >2
\]
so that subsequent sums may be written over fixed index ranges.

\subsection{ Fourier representation of the governing terms}
We now derive the Fourier representation of the  governing equation with the five-mode truncation and obtain the  corresponding reduced dynamical system for
the Fourier coefficients  $A_n(t)$ of the modes.

We begin with the linear operators, whose Fourier representations are immediate, and then derive the nonlinear interaction coefficients that organize the reduced dynamics into the underlying NLS four-wave interaction structure together with its higher-order and dissipative corrections.

\paragraph{Linear coefficients.}
Since the linear differential operators act diagonally on the Fourier basis, they do not generate
mode coupling.  For each Fourier mode $e^{in\mu x}$, the linear dispersive contribution is
\be
\Omega_n = \mu^2 n^2 - \frac{\epsilon}{2} \mu^3 n^3,
\ee
where the first term corresponds to the standard NLS dispersion and the second to the higher-order dispersive correction. The viscous damping contributes the diagonal modewise terms
\[
-i\Gamma A_n - 2i\epsilon\Gamma \mu n A_n,
\]
which act independently on each Fourier mode through a uniform and wavenumber-dependent decay. Together with the linear dispersive terms, these contributions do not alter the nonlinear coupling structure of the reduced system.
 All nontrivial  coupling of the modes, and hence the nonlinear energy transfer and phase-dependent interaction dynamics, arises through the nonlinear terms derived below.

\paragraph{Quadratic correlation coefficients:}
The nonlinear interaction terms are organized through the quadratic quantity
 $|u|^2$,
 whose Fourier coefficients generate the local cubic, mean-flow, and self-steepening contributions.  Substituting the truncated expansion \rf{five}
  into
\[
|u(x,t)|^2=u(x,t)\overline{u(x,t)},
\]
yields
\[
|u(x,t)|^2=\sum_{p=-4}^{4}Q_p(t)e^{ip\mu x},
\]
where
\[
Q_p
=
\sum_{\substack{k,\ell=-2\\ k-\ell=p}}^{2}
A_k\,\overline{A_\ell},
\qquad p=-4,\dots,4.
\]

Thus the quadratic term generates modulation harmonics extending over twice the original spectral support. The coefficients $Q_p$ encode the pairwise interaction products that organize the nonlinear interaction structure of the reduced system and form the common basis for the nonlinear coupling terms derived below.

Since  $|u|^2$ is real-valued, the coefficients satisfy the symmetry
\[
Q_{-p}=\overline{Q_p}.
\]

\paragraph{Cubic interaction coefficients:}
We next consider the local cubic nonlinearity $|u|^2u$. Substituting the expansions for $|u|^2$ and $u$, and collecting the coefficient of each retained Fourier mode $e^{in\mu x}$, gives

\[
|u|^2u
=
\sum_{n=-2}^{2}C_ne^{in\mu x},
\]
where
\[
C_n
=
\sum_{p=-4}^{4}Q_pA_{n-p}.
\]
Equivalently, using the definition of \(Q_p\),
\[
C_n
=
\sum_{k,\ell=-2}^{2}
A_k\overline{A_\ell}A_{n-k+\ell},
\qquad n=-2,\ldots,2,
\]
with the convention that \(A_j=0\) for \(|j|>2\).

These coefficients are related to the standard four-wave interaction structure inherited from the cubic NLS dynamics. The evolution of each mode is coupled through the quadratic correlation field $Q_p$, producing the nonlinear energy exchange through local cubic interactions.

\paragraph{Mean-flow interaction coefficients:}
The HONLS model includes the nonlocal mean-flow term $u \mathscr{H}((|u|^2)_x)$, which represents the nonlinear feedback of the induced mean flow on the wave envelope.

Differentiating the quadratic expansion gives:
\[
(|u|^2)_x
=
\sum_{p=-4}^{4} ip\mu\,Q_p\,e^{ip\mu x}.
\]
Adopting the following convention,
\[
\mathscr{H}(e^{ip\mu x})=i\,\operatorname{sgn}(p)e^{ip\mu x},
\]
we obtain
\[
\mathscr{H}\!\left((|u|^2)_x\right)
=
-\mu\sum_{p=-4}^{4}|p|\,Q_p\,e^{ip\mu x}.
\]
Multiplying by
$
u=\sum_{m=-2}^{2}A_me^{im\mu x},
$
and collecting  the retained Fourier mode contributions yields
\[
u\mathcal H\!\left((|u|^2)_x\right)
=
-\mu\sum_{n=-2}^{2}M_ne^{in\mu x},
\]
where
\[
M_n
=
\sum_{\substack{p=-4\\p\neq0}}^{4}
|p|Q_pA_{n-p}.
\]

Unlike the local cubic interaction coefficient $C_n$, the mean-flow coefficient $M_n$  weights each interaction by the modulation index $|p|$, so higher modulation harmonics receive greater weight in the mean-flow coupling.
Although the quadratic terms $Q_p$ determine the interaction structure, the nonlocal factor $|p|$ alters the relative contribution of the interaction terms to the evolution of the retained modes.

In the conservative HONLS, this term represents the nonlinear mean-flow feedback. In the NLD-HONLS, the factor $(1 + i\beta)$ introduces a dissipative correction acting on the  same carrier-sideband interaction terms. Thus mean-flow damping modifies an existing interaction mechanism rather than introducing independent mode-wise dissipation. 

\paragraph{Self-steepening coefficients.}

Finally, the self-steepening contribution takes the form
\be
|u|^2u_x
= i\mu
\sum_{n=-2}^{2}D_n e^{in\mu x},
\ee
 where 
\[
D_n
=
\sum_{\substack{p=-4\\ n-p\in[-2,2]}}^{4}
(n-p)\,Q_p\,A_{n-p},
\qquad n=-2,\dots,2.
\]

This term arises from the derivative acting on the interacting Fourier mode and weights  each interaction by that mode's wavenumber $n-p$. Unlike the cubic and mean-flow interaction terms, which preserve the same underlying interaction products $Q_p$,
the factor $n-p$ introduces a mode-dependent asymmetry into the nonlinear coupling, so that the same modal correlations contribute differently across the retained modes. 

Together, the coefficients $C_n$, $M_n$, and $D_n$ 
 define the  nonlinear interaction structure: the cubic term provides the fundamental carrier-sideband four-wave interaction, the mean-flow term reweights the same interaction structure through nonlocal modulation, and the self-steepening term introduces mode-dependent asymmetry into that exchange.

 \subsection{Projected five-mode system}

Substituting the Fourier coefficients derived  above into the governing equation \rf{DHONLS}  and  collecting terms for each  Fourier mode $e^{in\mu x}$, $n = -2, \dots, 2$,  yields the closed reduced system
\be
i\bdot A_n
=
\Omega_n A_n
-2C_n
+2\varepsilon\mu(1+i\beta)M_n
-8\varepsilon\mu D_n
-i\Gamma A_n
-2i\varepsilon\Gamma\mu n A_n
 : = F_n.
\label{F_neqn}
\ee

For clarity, we decompose $F_n$ as follows: 
\be
F_n = F_n^{\mathrm{NLS}} + F_n^{\mathrm{HO}} + F_n^{\mathrm{D}},
\ee
where
\begin{align}
F_n^{\mathrm{NLS}} &= \mu^2 n^2 A_n - 2C_n, \nonumber \\
F_n^{\mathrm{HO}} &= -\frac{\epsilon}{2}\mu^3 n^3 A_n + 2\epsilon \mu M_n - 8\epsilon \mu D_n, \\
F_n^{\mathrm{D}} &= 2i\epsilon \beta \mu M_n - i\Gamma A_n - 2i\epsilon \Gamma \mu n A_n. \nonumber
\end{align}

Here $F_n^{\mathrm{NLS}}$ governs the fundamental four-wave exchange underlying the modal energy transfer, while 
$F_n^{\mathrm{HO}}$ introduces the higher-order dispersive, mean-flow, and self-steepening corrections acting on the same inherited interaction structure.
The dissipative contribution $F_n^{\mathrm{D}}$ then modifies this structure either through modewise viscous decay or through the correlation-dependent mean-flow damping.

Together, these contributions define a closed reduced system for isolating how the conservative, higher-order, and dissipative terms enter the dominant
carrier-sideband interactions.

\subsubsection{Modal distinction between the dissipative mechanisms}

The reduced system \rf{F_neqn} makes explicit that the two dissipative models differ not in their underlying nonlinear interaction structure, but in the manner by which dissipation enters that structure.

In the viscous damping regime, $\beta = 0$, the dissipative contribution reduces to the diagonal terms
\be
-i\Gamma A_n - 2i\epsilon\Gamma\mu n A_n.
\label{gamma-term}
\ee
These terms act independently on each retained Fourier mode and therefore preserve the interaction structure generated by the cubic, mean-flow, and self-steepening terms. Although the conservative nonlinear terms continue to redistribute energy among the modes, the viscous contribution itself remains external to that interaction mechanism: it damps  each mode directly, with a wavenumber-dependent correction, but does not modify the underlying
interaction structure through which nonlinear energy transfer occurs.

By contrast, in the  nonlinear mean-flow damped regime, where $\Gamma =0$, the dissipative contribution is contained in the imaginary component of the mean-flow interaction term,
\be
2i\epsilon\beta\mu M_n.
\label{beta-term}
\ee
Here damping  acts through the same terms  $M_n$
 that govern the conservative mean-flow feedback. Since $M_n$ 
 is constructed from the quadratic correlation field $Q_p$,
 its contribution depends on the relative amplitudes and phases of the interacting modes.
 Thus, mean-flow damping modifies the same interaction structure responsible for the conservative carrier-sideband energy exchange rather than introducing independent modewise decay.
 
 The reduced system therefore distinguishes the two damping mechanisms at the modal level: viscous damping acts through independent modewise decay, whereas nonlinear mean-flow damping acts through the same coupled modal products that enter the conservative mean-flow interaction. Because these couplings depend on the relative amplitudes and phases of the interacting modes, the nonlinear mean-flow contribution retains an explicit phase dependence absent from the viscous terms.
 
 To make this phase dependence explicit, we next reformulate the reduced system in amplitude–phase variables and derive the corresponding interaction-phase dynamics.

\section{Amplitude--Phase Formulation and Interaction-Phase Dynamics}

To determine how the modal distinction identified in Section 3 enters the
phase combinations governing the carrier-sideband interactions, 
we let
\be
  A_n(t)=r_n(t)e^{i\phi_n(t)}, \qquad n=-2,\dots,2,
\ee
where $r_n(t)\geq 0$ and $\phi_n(t)\in\mathbb{R}$ denote the amplitude and phase of each mode.

Substituting this decomposition into the reduced system 
$i\bdot A_n = F_n$, where $F_n$ is given by \rf{F_neqn}, yields
\be
i\bdot r_n-r_n\bdot\phi_n=e^{-i\phi_n}F_n,
\ee
Separating real and imaginary parts gives the amplitude  equations
\be
\bdot r_n=\Im\!\left(e^{-i\phi_n}F_n\right),
\ee
and the phase equations
\be
r_n\bdot\phi_n=-\Re\!\left(e^{-i\phi_n}F_n\right).
\label{phase_eqn}
\ee

The amplitude equations govern the redistribution and dissipation of modal energy among the retained Fourier modes, while the phase equations determine
how the modal phases enter the nonlinear interactions.
The full amplitude-phase equations are recorded in Appendix A.
For the present analysis, the relevant variables are the specific phase combinations associated with the dominant carrier-sideband interactions.

The viscous damping terms \rf{gamma-term}
are purely imaginary after multiplication by \(e^{-i\phi_n}\) and therefore contribute only to the amplitude equations.

\paragraph{Interaction phases.}

To identify the phase variables governing the reduced dynamics,  observe that under the amplitude--phase decomposition
a typical  cubic term takes  the form

\be
A_k\overline{A_\ell}A_{n-k+\ell}
=
r_k r_\ell r_{n-k+\ell}
e^{i(\phi_k-\phi_\ell+\phi_{n-k+\ell})}.
\ee
Thus the nonlinear coupling depends not on the individual modal phases themselves, but on specific relative phase combinations determined by the underlying four-wave interactions.

This reflects the global phase-shift symmetry of the reduced system. Under the transformation
\[
A_n \mapsto A_n e^{i\theta},
\]
the phase combinations entering the nonlinear interactions are unchanged, so the dynamics depends on combinations of modal phases rather than on their absolute values.

The dominant four-wave exchanges involve the carrier mode together with the first and second symmetric sideband pairs. These interactions generate the principal phase combinations
\[
2\phi_0-\phi_{-1}-\phi_1  \qquad and \qquad  2\phi_0-\phi_{-2}-\phi_2.
\]
This  motivates the principal interaction phases
\be
\psi_1 = \phi_1+\phi_{-1}-2\phi_0,
\qquad
\psi_2 = \phi_2+\phi_{-2}-2\phi_0.
\label{psi_eqn}
\ee
Thus, $\psi_j$ measures the phase mismatch between the carrier and the jth symmetric sideband pair.
Equivalently, $\psi_1$ and $\psi_2$ arise directly as the arguments of the four-wave interaction products
\[
A_1A_{-1}\overline{A_0}^{\,2},
\qquad
A_2A_{-2}\overline{A_0}^{\,2}.
\]
The interaction phases therefore track the evolving phase mismatches of the principal carrier-sideband interactions.

Additional phase combinations arise through neighboring sideband interactions. In particular,
\be
\chi_+=\phi_2+\phi_0-2\phi_1,
\qquad
\chi_-=\phi_{-2}+\phi_0-2\phi_{-1},
\ee
correspond to interactions between adjacent sideband levels. Since these interactions involve smaller modal amplitudes, they enter  at higher order in the carrier-sideband hierarchy and are treated as secondary in the present reduction.
The interaction-phase dynamics therefore admits a natural hierarchy: the principal phases $\psi_1$ and $\psi_2$ govern the dominant carrier-sideband interactions, while the secondary phases $\chi_\pm$ contribute higher-order corrections.

\subsection{Exact interaction-phase equations}
Substituting the expression for $F_n$, \rf{F_neqn},  into the phase equations \rf{phase_eqn}
gives  
\be
r_n\bdot{\phi}_n
=
-\Omega_nr_n
+
2\Re\!\left(e^{-i\phi_n}C_n\right)
-
2\varepsilon\mu\Re\!\left(e^{-i\phi_n}M_n\right)
+
2\beta\mu\varepsilon
\Im\!\left(e^{-i\phi_n}M_n\right)
+
8\varepsilon\mu
\Re\!\left(e^{-i\phi_n}D_n\right).
\ee
The first term represents the linear dispersive phase mismatch contribution, while the remaining terms arise respectively from the cubic four-wave interactions, conservative mean-flow feedback, dissipative mean-flow corrections, and self-steepening interactions.

Substituting these expressions into the differentiated interaction-phase relations \rf{psi_eqn} and collecting terms gives
\begin{equation}
\bdot{\psi}_j
=
\Delta_j
+
\mathcal{C}_j
+
\mathcal{M}_j
+
\mathcal{D}_j,
\qquad
j = 1, 2
\label{4.9}
\end{equation}
Here the linear phase-mismatch terms are
\begin{equation}
\Delta_j
=
-\Omega_j-\Omega_{-j}+2\Omega_0,
\qquad
j = 1, 2
\label{4.13}
\end{equation}
The quantities
\(
\mathcal{C}_j,
\mathcal{M}_j,
\mathcal{D}_j
\)
denote the contributions of  the cubic, mean-flow, and self-steepening interaction terms, respectively, to the interaction-phase equations.
These contributions are obtained by combining the corresponding terms from
the individual phase equations when forming $\bdot \psi_j$. 
Explicit expressions are provided  in Appendix~A.

The interaction-phase equations therefore retain the same decomposition
into linear phase mismatch, nonlinear four-wave exchange, mean-flow feedback, and self-steepening corrections, now expressed directly in terms of the
phase combinations associated with the dominant carrier–sideband interactions.
In particular, the nonlinear mean-flow damping contribution enters through the term
\[
2\epsilon\mu\beta
\Im\!\left(
e^{-i\phi_n}M_n
\right),
\]
which contributes directly to the interaction-phase dynamics when
\(
\beta \neq 0
\).
In the viscous system, where $\beta = 0$,
this contribution is absent: viscous damping affects the interaction phases only indirectly through the evolving amplitudes of the modes.

\subsection{Carrier-sideband  approximation}
We now evaluate the interaction-phase equations under the amplitude hierarchy
\begin{equation}
r_0 \gg r_{\pm1} \gg r_{\pm2}.
\label{r_nscale}
\end{equation}

The purpose of this approximation is to identify the leading interaction-phase contribution generated by nonlinear mean-flow damping within the
principal carrier--sideband interaction regime.

From the exact interaction-phase equations
\begin{equation}
\bdot{\psi}_j
=
\Delta_j
+
\mathcal{C}_j
+
\mathcal{M}_j
+
\mathcal{D}_j,
\qquad
j=1,2,
\label{4.16}
\end{equation}
the distinguished interaction-phase contribution generated by nonlinear mean-flow damping arises from the
\(
2\epsilon\mu\beta
\)-dependent contribution involving the imaginary part of the mean-flow term,
\begin{equation}
2\epsilon\mu\beta
\Im\!\left(
e^{-i\phi_n}M_n
\right).
\label{4.17}
\end{equation}

For the first interaction phase \(
\psi_1
\), the dominant carrier--sideband interactions generate the leading terms
\begin{equation}
\Im\!\left(
e^{-i\phi_1}M_1
\right)
\sim
-r_0^2r_{-1}\sin(\psi_1),
\label{4.18}
\end{equation}

\begin{equation}
\Im\!\left(
e^{-i\phi_{-1}}M_{-1}
\right)
\sim
-r_0^2r_1\sin(\psi_1),
\label{4.19}
\end{equation}
and
\begin{equation}
\Im\!\left(
e^{-i\phi_0}M_0
\right)
\sim
2r_0r_1r_{-1}\sin(\psi_1).
\label{4.20}
\end{equation}
Substituting these leading contributions into the interaction-phase equation yields the dissipative mean-flow contribution
\begin{equation}
\mathcal{M}^{(\mathrm{diss})}_1
\sim
-\kappa_1\sin(\psi_1),
\label{4.21}
\end{equation}
where
\begin{equation}
\kappa_1
=
2\epsilon\mu\beta
\left[
r_0^2
\left(
\frac{r_{-1}}{r_1}
+
\frac{r_1}{r_{-1}}
\right)
+
4r_1r_{-1}
\right].
\label{4.22}
\end{equation}

Thus the nonlinear mean-flow damping contribution to the first principal interaction phase takes a restoring-type form with $\kappa_1$ determined by the carrier--sideband amplitudes.

An analogous calculation for the second principal interaction phase yields
\begin{equation}
\mathcal{M}^{(\mathrm{diss})}_2
\sim
-\kappa_2\sin(\psi_2),
\label{4.23}
\end{equation}
with
\begin{equation}
\kappa_2
=
2\epsilon\mu\beta
\left[
2r_0^2
\left(
\frac{r_{-2}}{r_2}
+
\frac{r_2}{r_{-2}}
\right)
+
8r_2r_{-2}
\right].
\label{4.24}
\end{equation}
For $\beta >0$ and nonzero sideband amplitudes, the coefficients
$\kappa_j$  are positive, $\kappa_j > 0$ , $j = 1,2$. 
The leading carrier-sideband reductions are provided in Appendix~A.

For the viscous system (\(
\beta=0
\)), the interaction-phase equations take the form
\begin{equation}
\bdot{\psi}_j
=
\Delta_j^{(V)}
+
\mathcal{C}_j^{(V)}
+
\mathcal M_j^{(V)}
+
\mathcal{D}_j^{(V)}
+
\mathcal{R}_j^{(V)},
\qquad
j=1,2,
\label{4.25}
\end{equation}
where $\Delta_j^{(V)}$, $\mathcal C_j^{(V)}$, $\mathcal M_j^{(V)}$, and $\mathcal D_j^{(V)}$ denote the corresponding leading-order phase-mismatch, cubic, conservative mean-flow, and self-steepening contributions under the carrier--sideband approximation, while $\mathcal R_j^{(V)}$ collects the remaining higher-order interaction terms arising within the five-mode interaction-phase equations.

Since viscous damping acts only through diagonal modewise decay, no direct restoring-type contribution appears in the interaction-phase equations.

On the other hand, for the nonlinear mean-flow damped system (\(
\Gamma=0,
\beta\neq0
\)) the interaction-phase equations become
\begin{equation}
\bdot{\psi}_j
=
\Delta_j^{(NLD)}
+
\mathcal{C}_j^{(NLD)}
+
\mathcal M_j^{(NLD)}
+
\mathcal{D}_j^{(NLD)}
-
\kappa_j\sin(\psi_j)
+
\mathcal{R}_j^{(NLD)},
\qquad
j=1,2,
\label{4.26}
\end{equation}
where the interaction-phase restoring contribution
\be
-\kappa_j\sin(\psi_j)
\label{restoring}
\ee
arises directly from the nonlinear mean-flow damping term, while the conservative mean-flow contributions remain contained in $\mathcal M_j^{(NLD)}$.

The carrier-sideband hierarchy \rf{r_nscale} introduced above is used
to identify the leading interaction-phase contributions associated with
the principal carrier--sideband four-wave interaction products. During strong focusing events, sideband amplitudes may become comparable to the carrier, and the hierarchy need not remain uniformly valid throughout the full PDE evolution.

\subsection{Interpretation of the reduced interaction-phase dynamics}

The interaction-phase equations reveal a fundamental difference between the two dissipative mechanisms.
In the viscous model, dissipation acts only through direct damping of the Fourier-mode amplitudes and does not contribute explicitly to the leading interaction-phase equations. The interaction-phase evolution is therefore governed by the conservative nonlinear interaction terms together with the indirect influence of viscous amplitude decay.

By contrast,  nonlinear mean-flow damping acts through the same coupled interaction terms that govern the conservative mean-flow feedback. At leading order, this produces the direct phase-dependent contribution \rf{restoring}
in each principal interaction-phase equation. In contrast with viscous damping, nonlinear mean-flow damping therefore acts directly within the carrier–sideband interaction structure.

The restoring character of this contribution is illustrated by the scalar equation
\[
\bdot\psi=-\kappa\sin(\psi),
\qquad
\kappa>0.
\]
 Near a phase-aligned state,
\[-\kappa_j\sin(\psi_j)\approx-\kappa_j\psi_j.\]
Thus, for $\kappa_j > 0$, the nonlinear damping contribution acts as
a restoring
feedback: when $\psi_j$ is displaced slightly to either side of the phase-aligned state, the contribution has the opposite sign and tends to drive  $\psi_j$ back towards  that state.

In the full reduced equations, however, this contribution competes with the dispersive, cubic, conservative mean-flow, self-steepening, and higher-order interaction terms. Consequently, the presence of $-\kappa_j\sin(\psi_j)$ does not by itself imply phase locking,
bounded interaction-phase evolution, or persistence of any particular Floquet spectral configuration.

The interaction phases may therefore remain dynamically active and evolve asymmetrically even in a recurrent regime; the restoring
 contribution modifies their evolution without requiring them to become stationary or bounded. This observation motivates the benchmark analysis of Section 5, where substantial and unequal interaction-phase evolution is shown to coexist with recurrent modulation and an unchanged Floquet spectral configuration.

The five-mode reduction is therefore used as a mechanism-identification model rather than as a quantitatively complete description of the full PDE dynamics or its nonlinear Floquet spectrum. In Section 5, the interaction phases are computed directly from the full PDE solutions and interpreted together with recurrent focusing and the independently established Floquet spectral regimes.

\section{Numerical interaction-phase diagnostics and modulation dynamics}

We now compute the principal interaction phases directly from the full PDE solutions and examine their evolution together with recurrent focusing and the Floquet spectral classifications summarized in Section 2. Finite-gap NLS solutions are first used as benchmarks for interpreting interaction-phase evolution within known recurrent spectral states; the same diagnostics are then applied to the
NLD-HONLS and V-HONLS systems.
\subsection{Numerical framework}

The NLD-HONLS and V-HONLS equations are numerically integrated using
a Fourier pseudo-spectral discretization in space and a fourth-order exponential time-differencing Runge-Kutta method \cite{CM2002}.
 The spatial resolution and time step  are selected according to the complexity of the evolving solution.
For the simulations considered here, the spatial period is   $L = 4\sqrt{2} \pi$, with  $N = 256$  Fourier modes  and  time step 
$\Delta t = 10^{-3}$.

To assess numerical accuracy, we initialize the conservative HONLS equation
($\Gamma = 0, \beta = 0$, with $\epsilon \neq 0$), using the benchmark finite-gap NLS initial data given in Appendix B, evaluated at
$t = 0$. We monitor the conserved quantities of the HONLS system: the total energy E(t), momentum P(t), and Hamiltonian,
\be
E(t) = \frac{1}{L} \int_0^L |u|^2\,\mathrm{d}x, \qquad
P(t) = \frac{i}{2L}\int_0^L (uu^*_x -u^*u_x)\,\mathrm{d}x,
\label{EandPdefn}
\ee
and
\be\label{hamiltonian}
H(t) =  \int_0^L \left\{ | u_x |^2 - | u |^4 - \ri\epsilon\left[
1/4\left( u_x u_{xx}^* - u_x^* u_{xx} \right) 
+ 2 | u |^2\left(u^* u_x - u u_x^*\right) - |u|^2
\mathcal{H} \left( | u |^2_x \right)\right]\right\}\,dx.
\ee
Over the interval $0 \le t \le 100$, 
these quantities are  preserved to within $\mathcal{O} (10^{-11})$,
providing a stringent check on the numerical accuracy of the PDE integration used to compute the interaction-phase diagnostics.
The two dissipative  systems  are compared using the common SPB initial conditions given in \rf{SPB_ic}.

\subsubsection{Fourier-Energy Assessment of the Five-Mode Truncation}

To assess the range over which the five-mode reduction remains representative of the dominant carrier-sideband dynamics, we monitor the fraction of the full-PDE Fourier energy contained in the retained modes.
Define
\begin{equation}
\eta_{2K+1}(t)
=
\frac{
\displaystyle \sum_{n=-K}^{K}|A_n(t)|^2
}{
\displaystyle \sum_n |A_n(t)|^2
},
\qquad
K=0,1,2,\ldots.
\end{equation}

Thus, \(\eta_5\) measures the fraction of the total Fourier energy contained in the five modes retained by the reduced system, while \(\eta_9\)
measures the corresponding fraction for larger carrier-centered mode sets.
By construction, $ 0 \leq \eta_{2K+1} \leq 1$, with values near unity indicating that the selected carrier-centered modes contain nearly all of the resolved Fourier energy.

These quantities provide a direct measure of how much of the full-PDE Fourier content is represented by the reduced carrier-sideband description.
In particular, \(\eta_5\) assesses the instantaneous energy representation of the five-mode truncation, whereas comparison with \(\eta_9\) indicates whether energy omitted from the five-mode system remains concentrated in nearby Fourier harmonics or is distributed more broadly across the resolved spectrum.

\subsection{Interaction-phase diagnostics}

Writing the full PDE solution in Fourier form as
\[
u(x,t)=\sum_{n\in\mathbb{Z}} A_n(t)e^{in\mu x},
\]
 we define the principal carrier-centered interaction products by
\[
\mathcal{I}_j(t)
=
A_j(t)A_{-j}(t)\overline{A_0(t)}^{\,2}, \quad j = 1,2
\]
and the corresponding interaction phases 
\[
\psi_j(t)=\arg\!\left(\mathcal{I}_j(t) = \phi_j+\phi_{-j}-2\phi_0,\right).
\]

These are the same carrier-sideband interaction phases introduced in Section 4, but are evaluated here directly from the Fourier coefficients of the full PDE solution rather than from the five-mode truncation. To distinguish genuine cumulative phase evolution from branch discontinuities associated with the principal argument, the interaction phases are continuously unwrapped in time after computing $\mathcal{I}_j(t)$.

To monitor the applicability of these carrier-centered diagnostics, we define the  Fourier spectral peak
\be
k_{peak}(t) = \mbox{argmax}_{k} |A_k(t)|.
\label{kpeak}
\ee
as the index  of the Fourier mode of  maximal amplitude at time $t$.
The quantity  $k_{peak}(t)$ is used  to verify that the dominant Fourier component remains sufficiently close to
the original carrier for $\psi_1$ and $\psi_2$
to remain meaningful diagnostics of the principal carrier–sideband interactions.
Moderate downshifting does not invalidate this interpretation because the interaction phases depend on relative mode separations; substantial migration of the dominant Fourier component, however, would. As a result, the dissipative comparisons below are restricted to intermediate times for which $k_{peak}$
 remains near the original carrier.
The quantity $k_{peak}$ is not a diagnostic of the nonlinear Floquet spectrum or of Floquet-band localization.

Focusing events marked in the benchmark figures are identified directly using the rogue-wave strength criterion
\rf{rwcondb}. For the dissipative SPB evolutions, the marked event times are those established in the prior Floquet
study \cite{SI2025}. The benchmark solutions may exhibit additional focusing events that do not satisfy the rogue-wave threshold and are therefore not marked. In both cases, the markers are used only as reference times for comparison with the interaction-phase, Fourier-energy, and Floquet dynamics.

\subsection{Interaction-phase benchmarks}

We first examine quasiperiodic NLS benchmark solutions whose modulation dynamics and Floquet spectral structure are independently known. These solutions provide reference cases  for interpreting interaction-phase evolution in settings where the underlying Floquet configuration does not change.

We consider two representative benchmark  solutions whose observable modulation dynamics are dominated by one or two nonlinear modulation modes, respectively.
The one-mode benchmark is initialized by evaluating the exact three-phase finite-gap solution (B.1) at $t = 0$. The two-mode benchmark is initialized directly from the perturbed data (B.2), whose subsequent evolution is well
approximated by  a five-phase finite-gap state, as discussed in Appendix~B. The labels ``one-mode'' and ``two-mode'' refer to the dominant nonlinear modulation structure rather than to the number of nonzero Fourier components. 
\begin{figure}[htp!]\centerline{
\includegraphics[width=.3\textwidth]{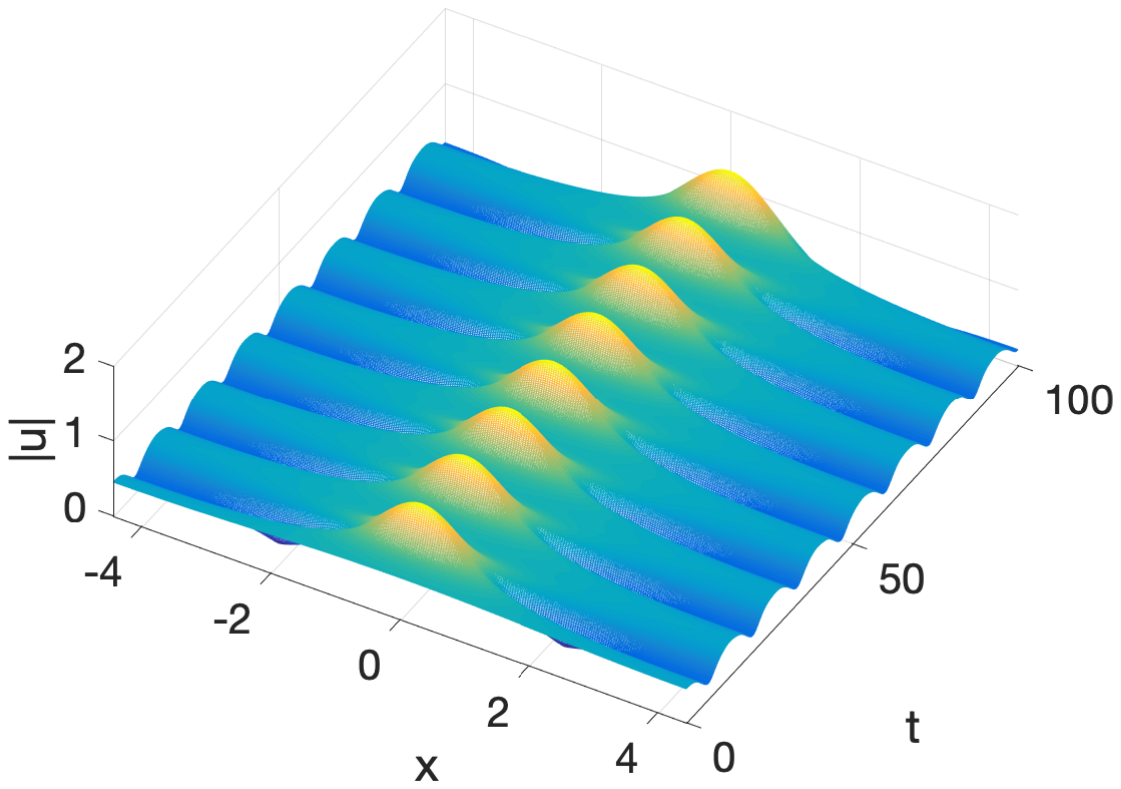}
\includegraphics[width=.3\textwidth]{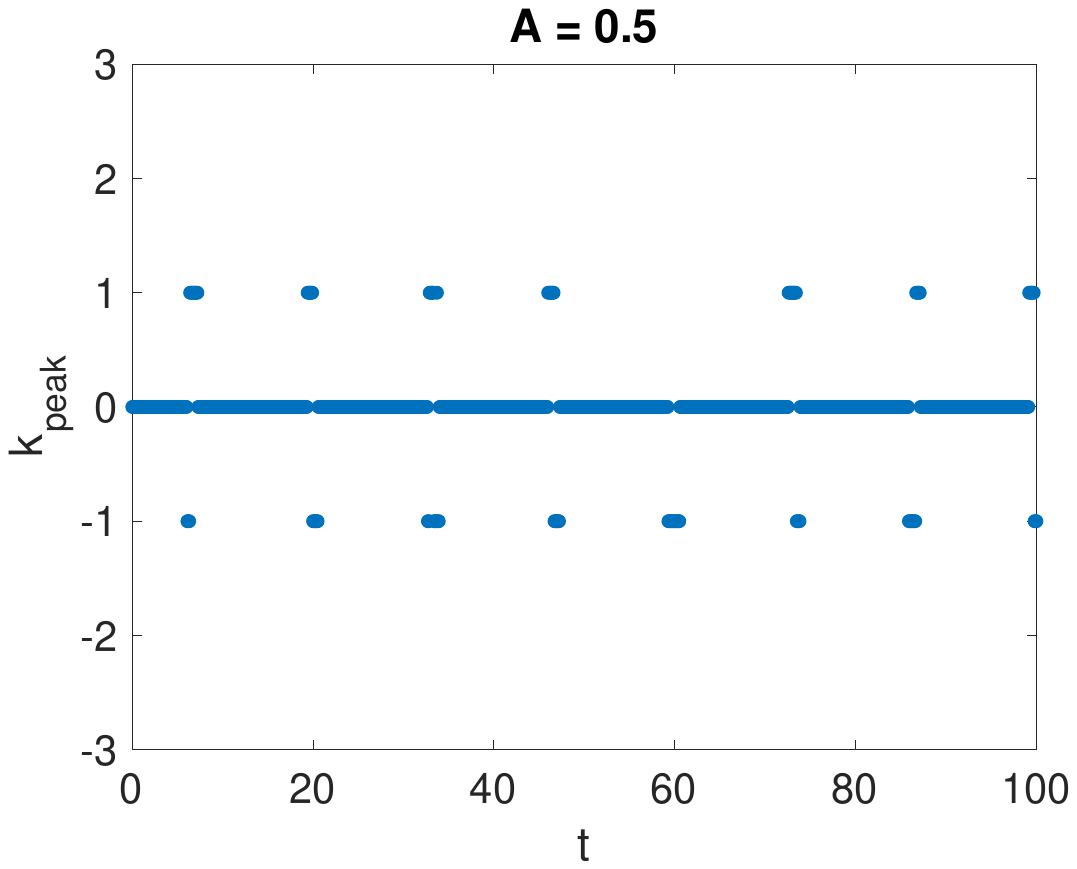}
\includegraphics[width=.3\textwidth]{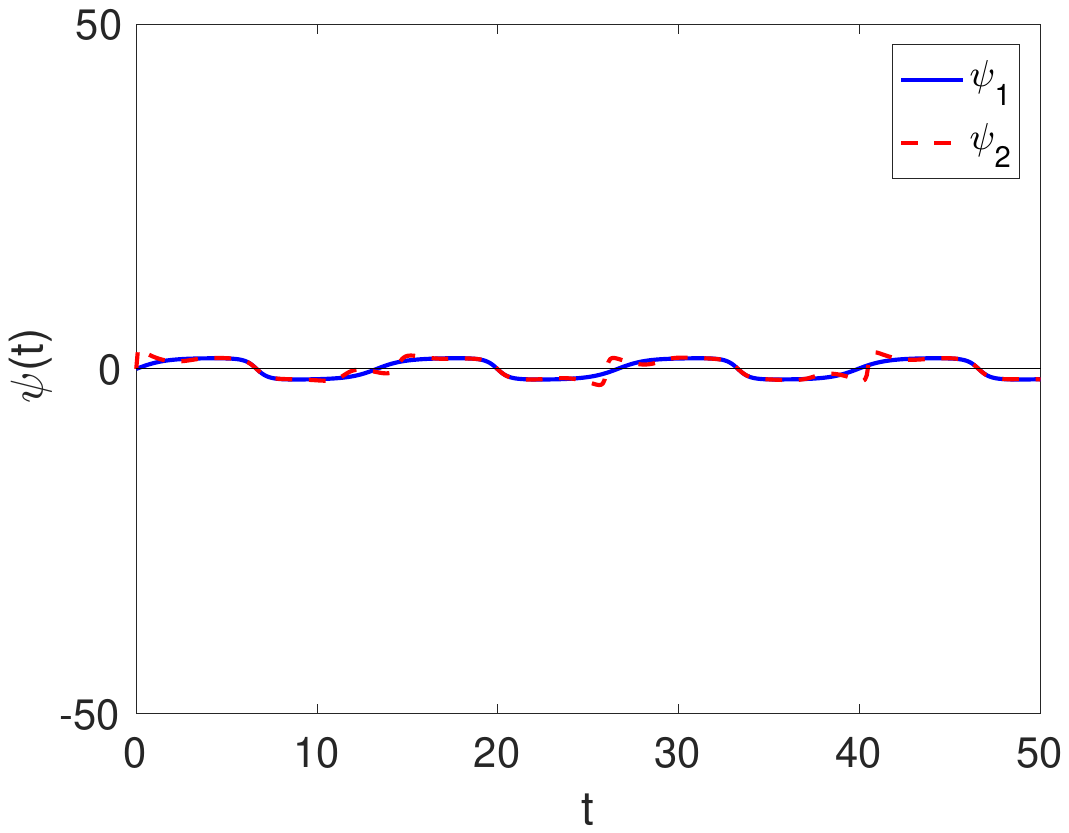}
}
    \vspace{-48pt}
  \centerline{(a)\hspace{0.33\textwidth}(b)\hspace{0.33\textwidth}(c)}
\caption{
  One-mode NLS benchmark: (a) surface evolution, (b) Fourier spectral peak
  $k_{peak}$, and (c) interaction phases $\psi_1$ and $\psi_2$, exhibiting recurrent modulation cycles and bounded phase evolution.
}
\label{fig:1UM}
\end{figure}

Figure~\rf{fig:1UM} shows the one-mode finite-gap dynamics. The surface evolution exhibits strongly recurrent focusing-defocusing  cycles while 
$k_{peak}$ remains close to the carrier throughout the evolution.
 Correspondingly, both interaction phases $\psi_1$ and $\psi_2$ 
 remain bounded and oscillatory, repeatedly returning to neighborhoods of their initial values during successive modulation cycles. This provides a simple reference case in which recurrent focusing  and recurrent interaction-phase evolution coexist with an invariant finite-gap structure.
\begin{figure}[htp!]
\centerline{
\includegraphics[width=.3\textwidth]{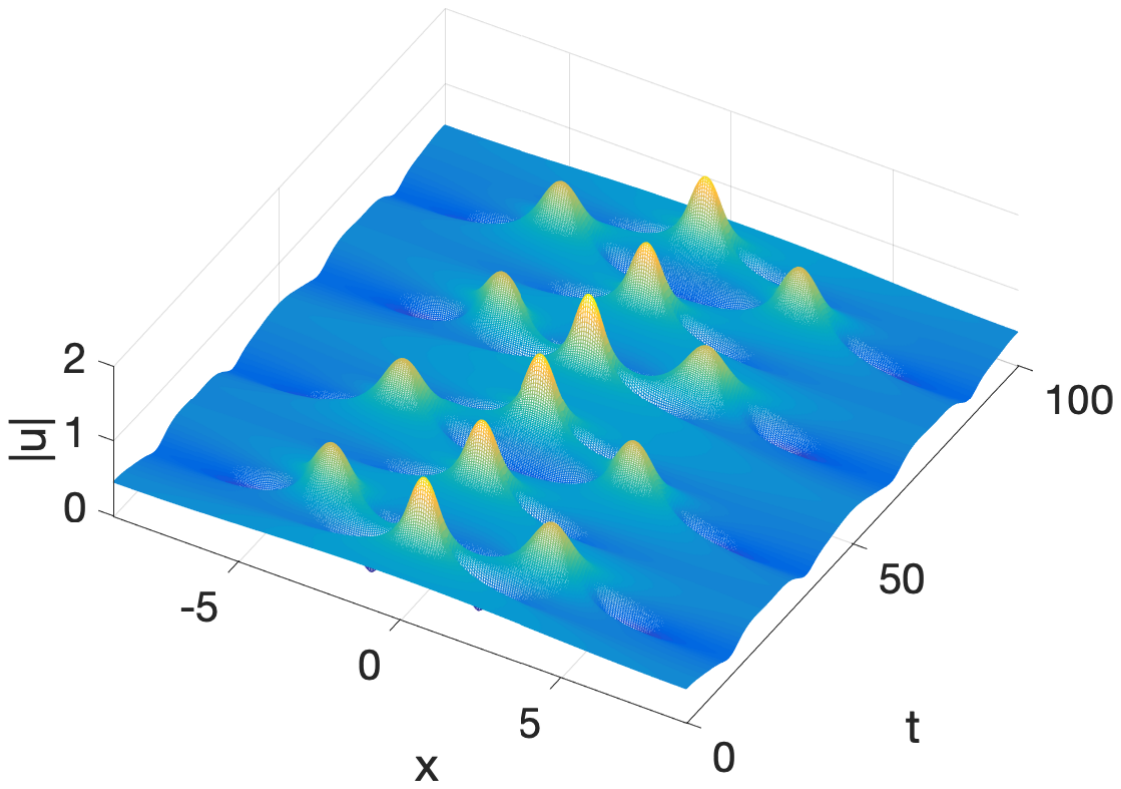}
\includegraphics[width=.3\textwidth]{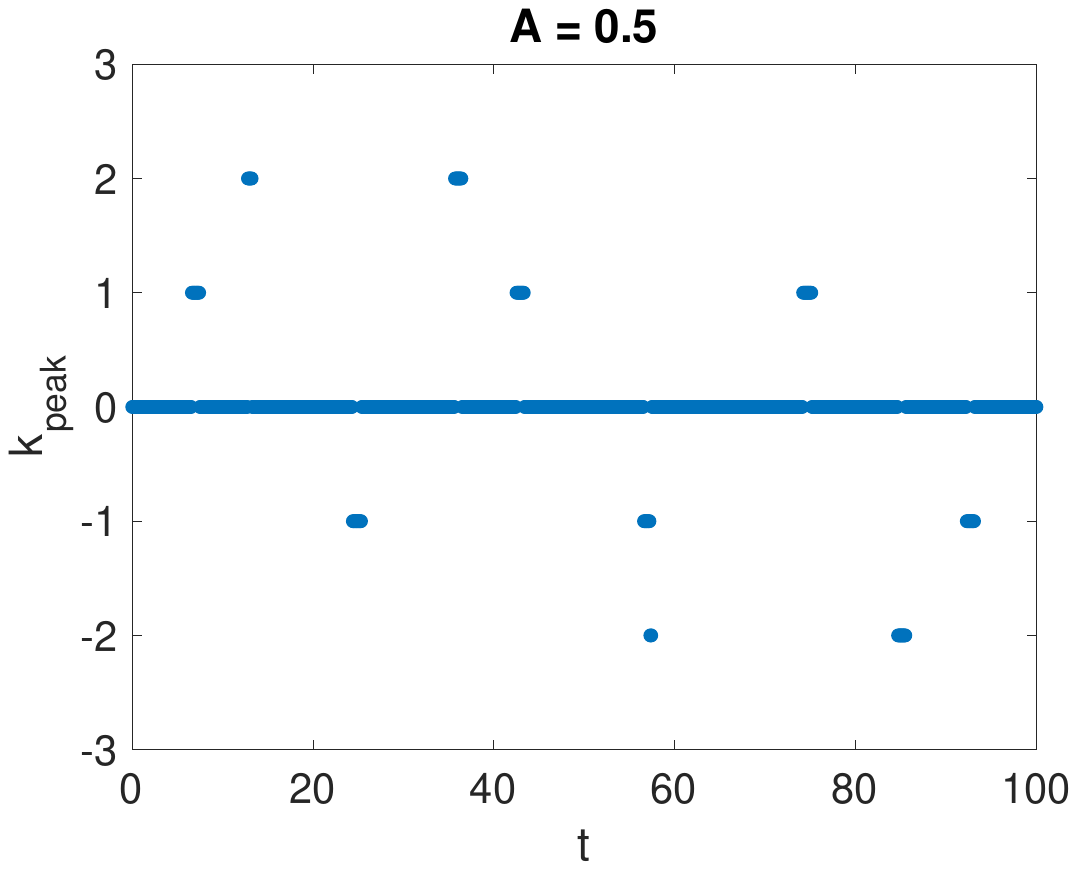}
\includegraphics[width=.3\textwidth]{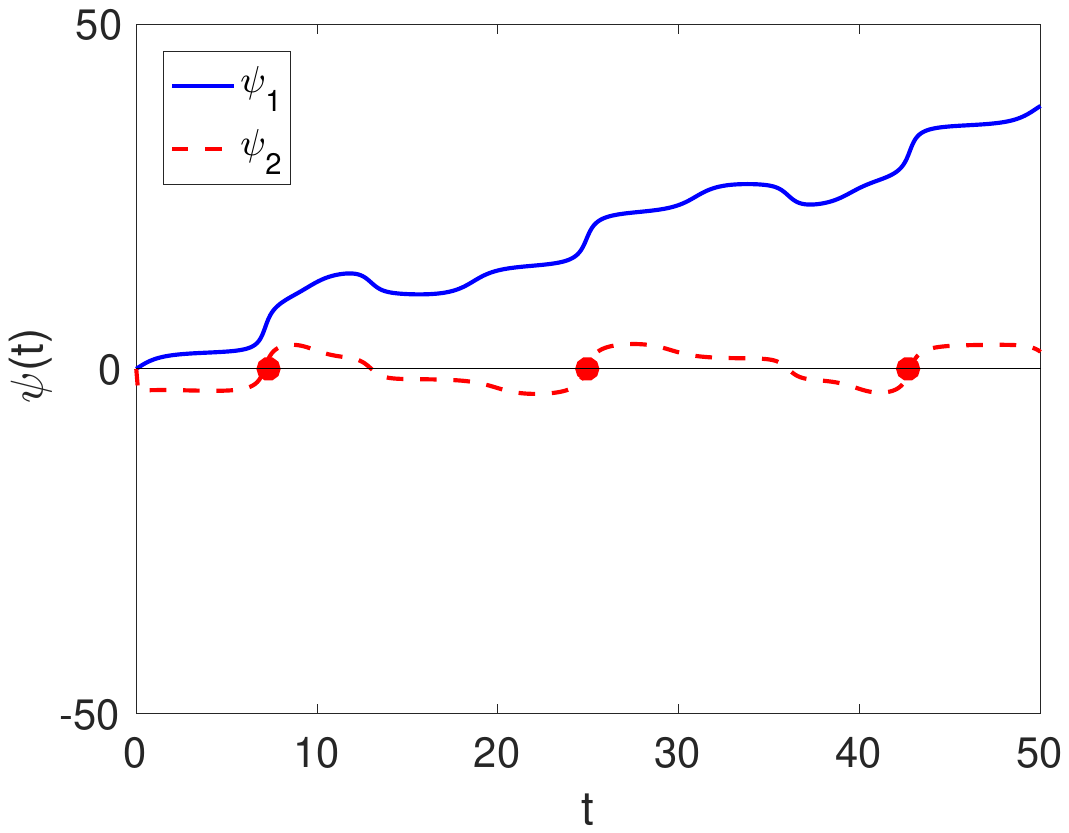}
}
    \vspace{-48pt}
  \centerline{(a)\hspace{0.33\textwidth}(b)\hspace{0.33\textwidth}(c)}
\caption{
  Two-mode  NLS benchmark: (a) surface evolution, (b) Fourier spectral peak $k_{peak}$, and 
  (c) interaction phases, $\psi_1$ and $\psi_2$ exhibiting recurrent modulation together with substantial cumulative phase drift. Red bullets indicate the
  strong focusing events satisfying the rogue-wave strength criterion \rf{rwcondb}.
}
\label{fig:2UM}
\end{figure}

In contrast, the two-mode finite-gap solution shown in Figure~\rf{fig:2UM} exhibits substantially richer quasiperiodic dynamics. The surface evolution
continues to  exhibit strongly recurrent focusing-defocusing behavior throughout the interval shown, while $k_{peak}$ 
remains close to the carrier.
The interaction phases differ qualitatively from
the one-mode benchmark.

In particular,   Figure~\rf{fig:2UM}(c) shows that  $\psi_2$  remains comparatively bounded and recurrent throughout the evolution, while  $\psi_1$ 
undergoes substantial cumulative drift over successive modulation cycles.
Nevertheless, both interaction phases retain the same overall direction of evolution. The red bullets mark the focusing events that satisfy the rogue-wave strength criterion \rf{rwcondb}.
The sharper changes in the slope and local oscillatory structure
of $\psi_1$ occur primarily 
near these marked focusing events, separating broader intervals of comparatively gradual evolution.

Importantly, these larger changes in the phase evolution occur near recurrent
strong focusing events  rather than during a breakdown of the quasiperiodic modulation dynamics. The two-mode benchmark suggests that the relevant distinction is not whether the interaction phases remain bounded, but whether their
larger changes remain tied to  the recurrent focusing  dynamics. Substantial cumulative phase drift and markedly different evolution of $\psi_1$ and $\psi_2$ can therefore coexist with recurrent modulation and an unchanged Floquet spectral configuration. The two-mode benchmark therefore provides a useful reference for interpreting the more complicated interaction-phase evolution of the dissipative SPB solutions considered next.

\subsection{Dissipative interaction-phase dynamics}

The Floquet spectral regimes considered below were established in \cite{SI2025} and summarized in Section~2. Under nonlinear mean-flow damping, the Floquet-band configuration evolves without critical-point crossings, whereas the viscous evolution undergoes repeated critical-point crossings and band reconfiguration.
The same SPB initial data and previously identified focusing events are used here to examine the corresponding full-PDE surface evolution, Fourier diagnostics, and interaction phases. Guided by the finite-gap benchmarks of Section~5.3, we focus in particular on whether the larger interaction-phase changes remain tied to recurrent focusing and whether
$psi_1$ and $\psi_2$ retain a common large-scale direction of evolution.

The Fourier diagnostics characterize the concentration of the solution about the carrier and assess the range over which the five-mode reduction remains representative of the dominant carrier-sideband dynamics, while the interaction phases characterize the evolving phase relationships among those interactions.
These quantities are therefore used to interpret the dynamics within the established Floquet regimes rather than to determine the spectral
classification itself.

The Fourier-energy fractions are shown over $0 \leq t \leq 40$, corresponding approximately to the interval before one of the dominant NLD-HONLS Floquet bands begins to exceed the spectral-localization threshold. This choice concerns the regime over which the reduced carrier-sideband description is being assessed. The interaction phases, by contrast, are computed directly from the full PDE and are shown over the longer interval
$0 \leq t \leq 100$ to examine their relation to recurrent focusing and the established Floquet evolution.
The $k_{peak}$
diagnostic is likewise shown over this longer interval to verify that
the dominant Fourier component remains near the original carrier.

Consistent with the common short-time Floquet evolution summarized in
Section~2.3,  the two dissipative systems also exhibit essentially the same initial focusing dynamics and similar interaction-phase evolution through the first rogue-wave event near $t \approx 3.5$. Their subsequent behavior diverges as the Floquet evolutions separate into the persistent regime for NLD-HONLS and the repeatedly reconfiguring regime for V-HONLS, providing the setting in which the benchmark distinctions identified in Section~5.3 become useful.

\begin{figure}[htp!]
    \vspace{-48pt}
\centerline{
\includegraphics[width=.35\textwidth]{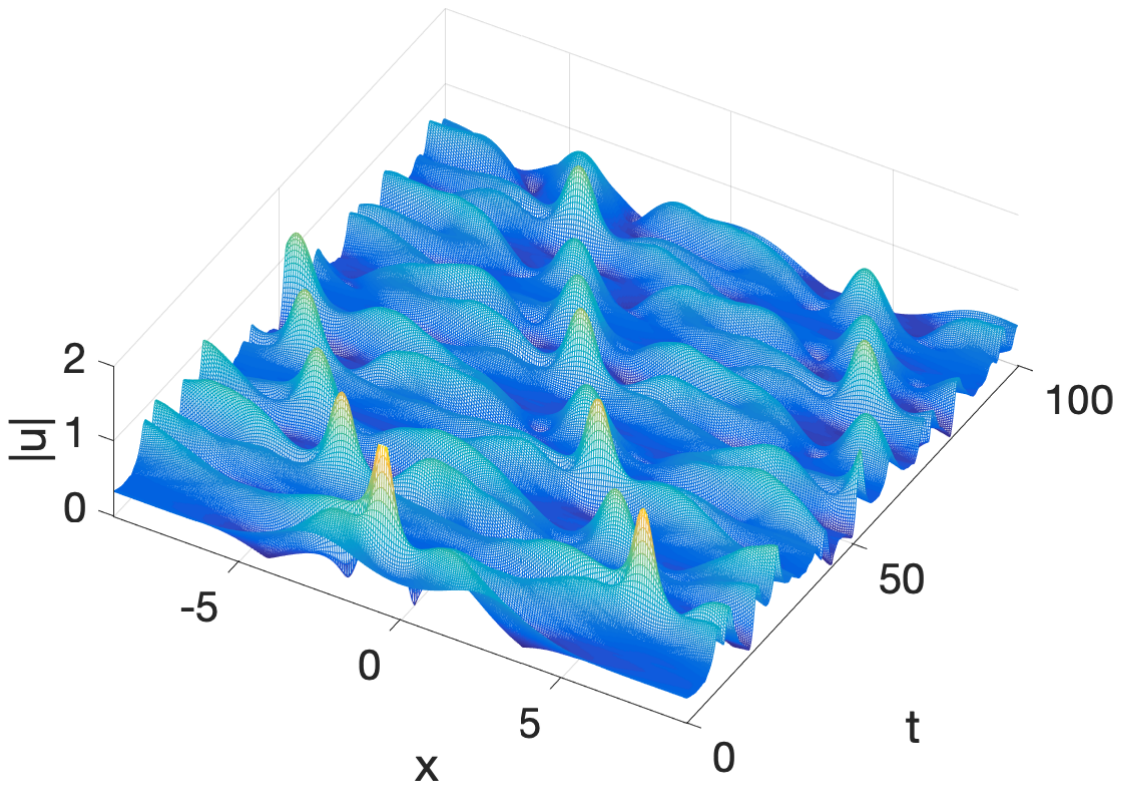}
    \includegraphics[width=.35\textwidth]{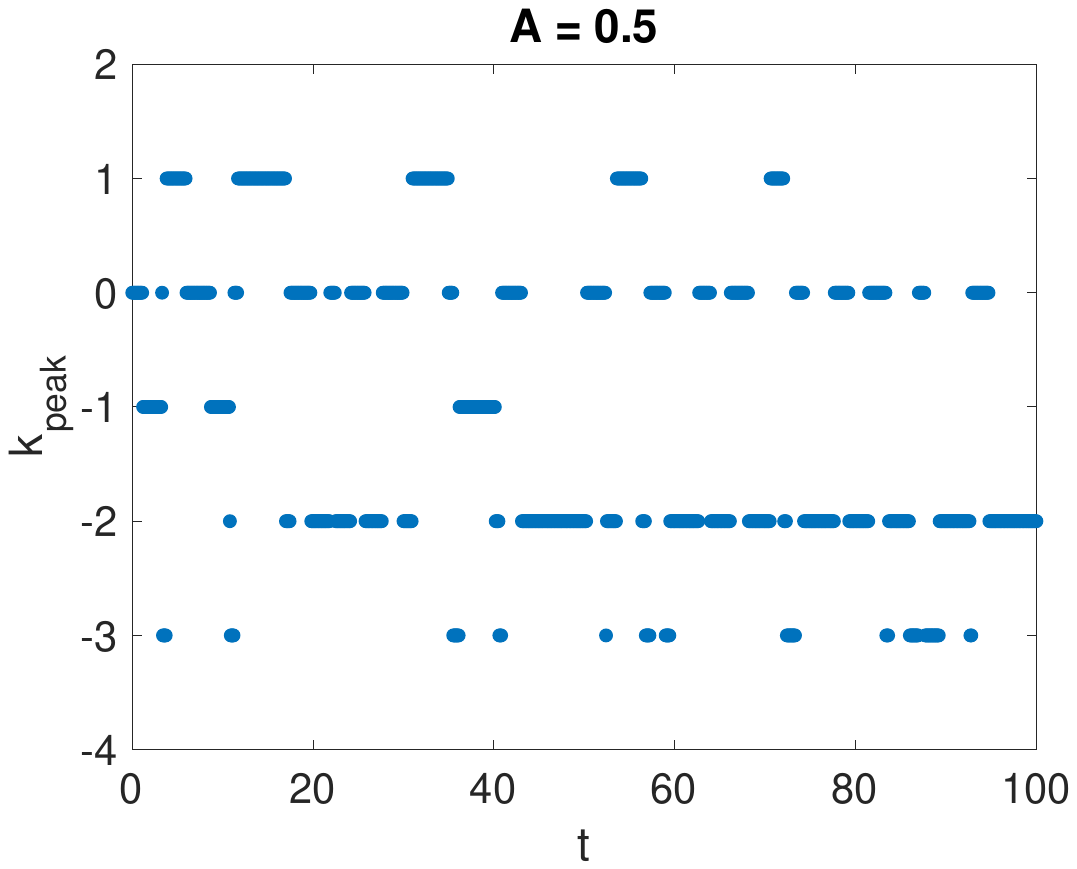}
    \includegraphics[width=.35\textwidth]{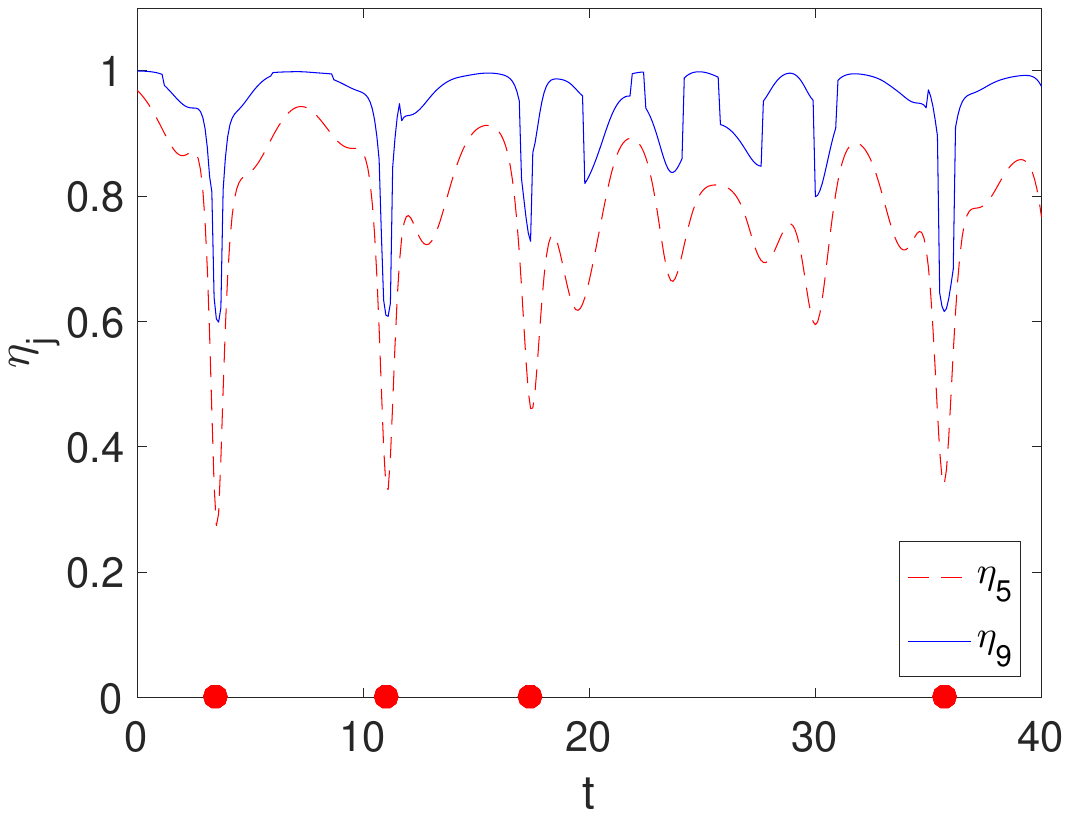}
}
    \vspace{-48pt}
  \centerline{(a)\hspace{0.33\textwidth}(b)\hspace{0.33\textwidth}(c)}
  \caption{NLD-HONLS dynamics with SPB initial data and $\beta = 0.1$: (a) surface evolution, (b) Fourier spectral peak $k_{peak}$, and (c) carrier-centered Fourier-energy fractions $\eta_5$ (dashed) and $\eta_9$ (solid).
Red bullets indicate the strong focusing events defined in Section~5.2. The larger $\eta_9$ 
 values show that energy leaving the central five modes remains concentrated primarily in nearby Fourier harmonics.}
\label{fig:nld_surf_kp_psi}
\end{figure}

For the NLD-HONLS system with $\beta = 0.1$,  the  surface evolution shown in Figure~\rf{fig:nld_surf_kp_psi}(a) exhibits recurrent localized focusing together with gradual frequency downshifting. The Fourier spectral peak $k_{peak}$, shown  in
Figure~\rf{fig:nld_surf_kp_psi}(b), remains close to the original carrier mode over the interval considered, supporting the continued interpretation of
$\psi_1$ and $\psi_2$ 
 as the principal carrier-centered interaction phases.
 The carrier-centered energy fractions in Figure~\rf{fig:nld_surf_kp_psi}(c) provide a complementary measure of the Fourier concentration.
 Away from the strongest focusing events, $\eta_5$ typically lies between approximately $0.65$ and $0.95$, while $\eta_9$ remains between approximately $0.85$ and $1$. 
The sharpest reductions in $\eta_5$
occur near the recurrent two-soliton-like rogue-wave focusing events identified
in \cite{SI2025} and marked by red bullets, indicating increased excitation of additional Fourier harmonics during strong localization. The corresponding $\eta_9$ (solid curve)
 values remain substantially larger, showing that much of the energy omitted from the five-mode truncation remains concentrated in nearby carrier-centered harmonics rather than spreading broadly across the resolved Fourier spectrum.
Thus, the strongest focusing events are accompanied by a temporary redistribution of Fourier energy beyond the central five modes, while the dominant energy remains concentrated in a relatively narrow carrier-centered set.

 Following the initial focusing event, the NLD-HONLS interaction phases develop a comparatively persistent large-scale evolution. As shown in Figure~\rf{fig:ble}(a),
 $\psi_1$  remains within a comparatively narrow range, while $\psi_2$ 
 undergoes greater cumulative drift.
 Despite this difference, both phases retain predominantly the same long-time drift direction.

 The sharper, short-time changes in slope and local oscillatory structure associated with individual focusing events occur primarily near the strong-focusing events marked by red bullets. These local changes are more pronounced in
 $\psi_1$,
 while the intervals between focusing events exhibit more gradual evolution.
 Thus, $\psi_1$
 shows the stronger local response to focusing even though $\psi_2$ 
 accumulates the greater long-time drift.

 This behavior is qualitatively similar to the two-mode benchmark of
 Section~5.3 in that the more pronounced local changes in $\psi_1$ are
 likewise tied to recurrent strong focusing. The principal difference
  is in the  cumulative drift:  $\psi_1$ 
 drifts more strongly in the benchmark, whereas $\psi_2$ 
 does so under NLD-HONLS.

 The corresponding Floquet evolution remains within the persistent spectral regime summarized in Section 2. Figure~\rf{fig:ble}(b) shows the lengths of the two dominant Floquet bands. Both initially satisfy the Floquet spectral-localization criterion (2.9); after approximately $t \approx 35-40$, one band gradually exceeds the localization threshold while the other remains localized.
 This loss of localization of one band does not constitute spectral reconfiguration: the dominant bands expand and drift while the Floquet spectrum remains free of critical-point crossings. Thus, the recurrent strong-focusing events and the associated local interaction-phase changes occur within a Floquet configuration that remains organized throughout the evolution.
\begin{figure}[htp!]
\centerline{
    \includegraphics[width=.33\textwidth]{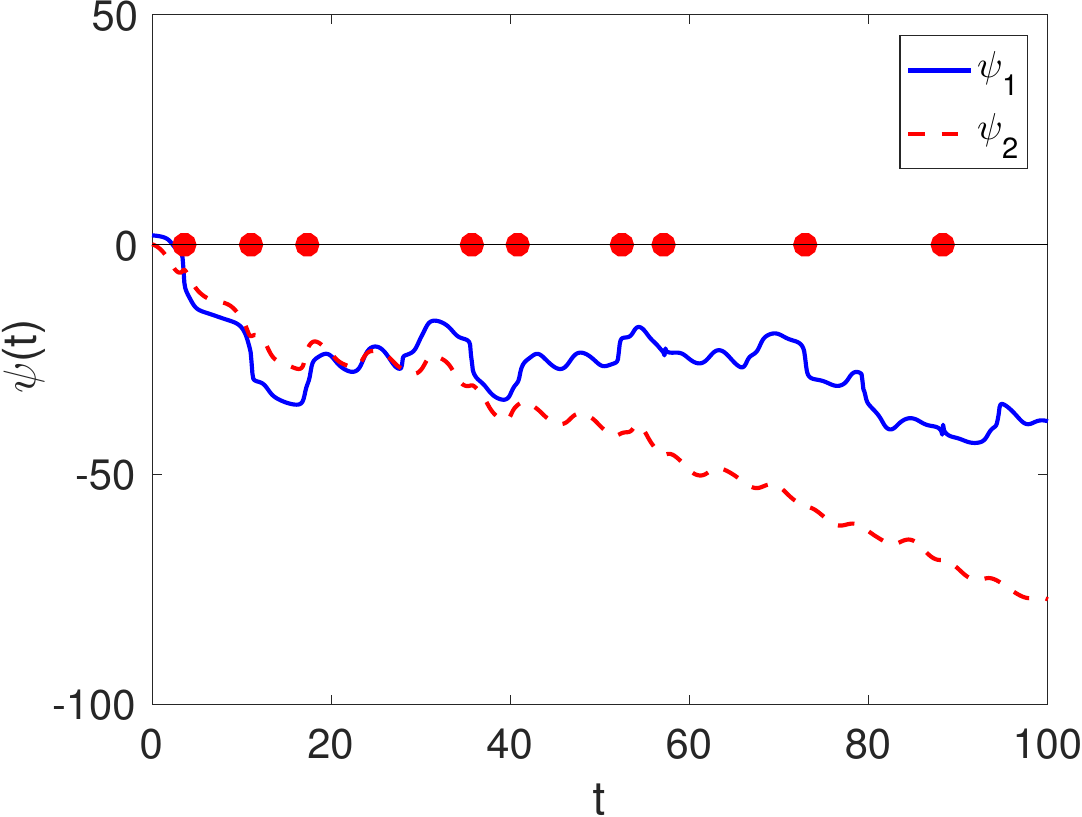}
\hspace{6pt}\includegraphics[width=0.33\textwidth]{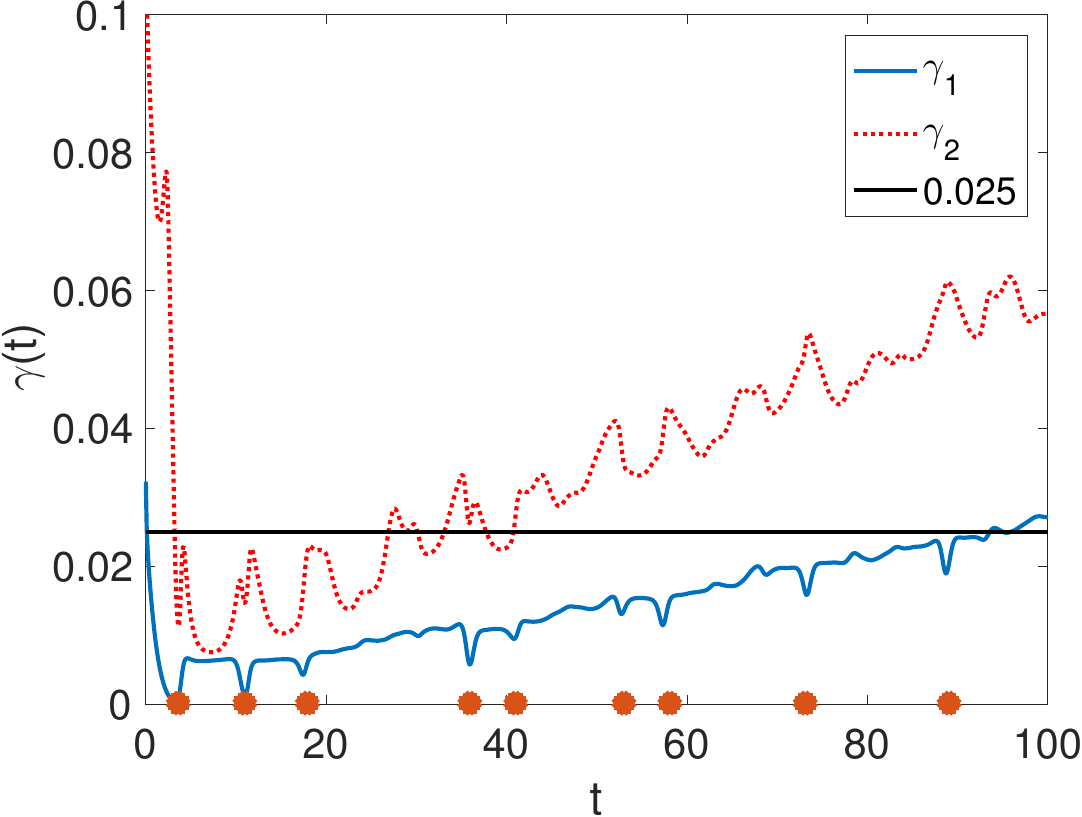}
}
  \centerline{(a)\hspace{0.33\textwidth}(b)}
\caption{NLD-HONLS interaction-phase and Floquet-band dynamics for \(\beta=0.1\): (a) principal interaction phases \(\psi_1\) and \(\psi_2\); (b) lengths of the two dominant Floquet bands, \(|\gamma_1(t)|\) and \(|\gamma_2(t)|\) with  the horizontal line denoting the Floquet spectral-localization threshold \(|\gamma|=0.025\). Red bullets in both panels indicate focusing events identified as described in Section~5.2.
  The first strong-focusing event near $t = 3.5$ coincides with the initial sharp decrease in $\psi_1$, 
and subsequent pronounced local changes in $\psi_2$
occur primarily near the marked focusing events.
Although one band exceeds the localization threshold after approximately \(t\approx 35\text{--}40\), the Floquet-band configuration remains free of critical-point crossings throughout the evolution.}
 \label{fig:ble}
\end{figure}

By contrast, for the V-HONLS system with $\Gamma = 0.002$, the surface evolution in Figure~\rf{fig:visc9}(a)  becomes  progressively
less recurrent, although intermittent focusing events continue to occur.
The Fourier spectral peak  $k_{peak}$, shown  in
Figure~\rf{fig:visc9}(b), remains sufficiently close to the original carrier over the interval considered for
$\psi_1$ and $\psi_2$ to remain meaningful carrier-centered diagnostics.
The five-mode energy fraction $\eta_5$,  shown in Figure~\rf{fig:visc9}(c), also remains comparatively large over most of the interval,
with temporary decreases near strong-focusing events followed by recovery to values near unity.
Thus, the repeated Floquet spectral reconfigurations of the viscous solution cannot be attributed simply to a progressive loss of Fourier concentration in the central carrier-sideband modes.
\begin{figure}[htp!]
  \centerline{
  \includegraphics[width=.35\textwidth]{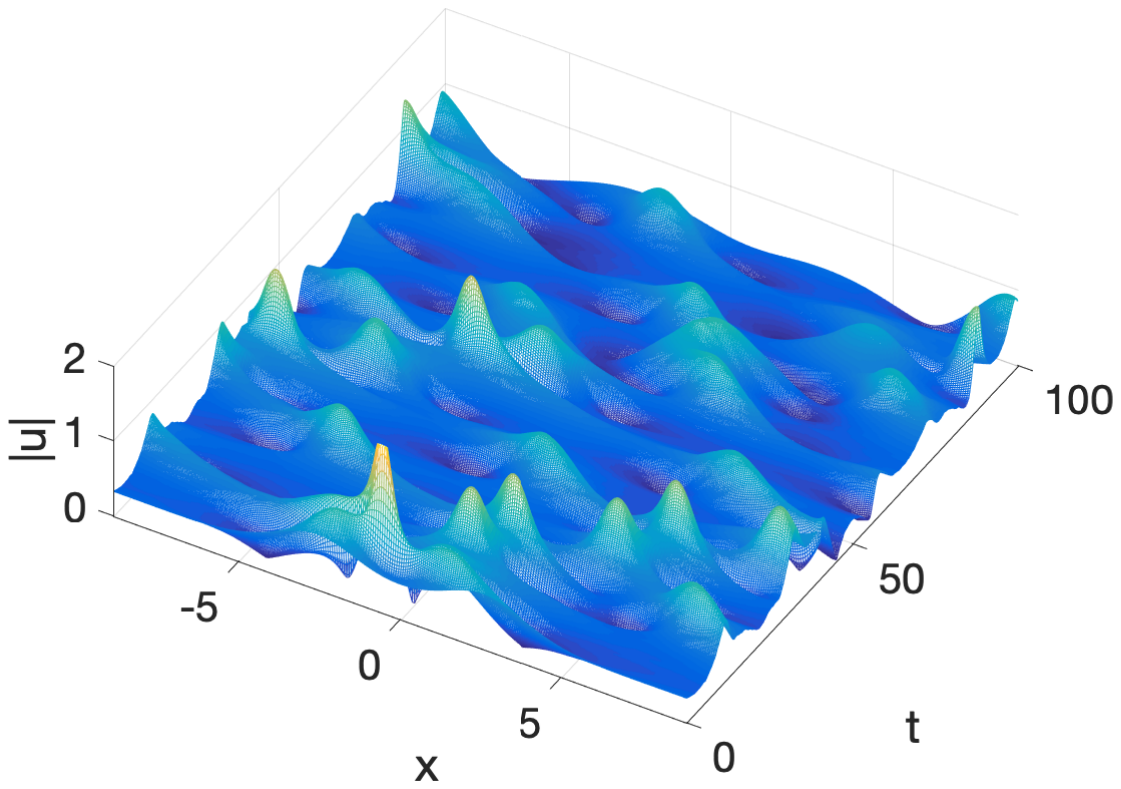}
\includegraphics[width=.35\textwidth]{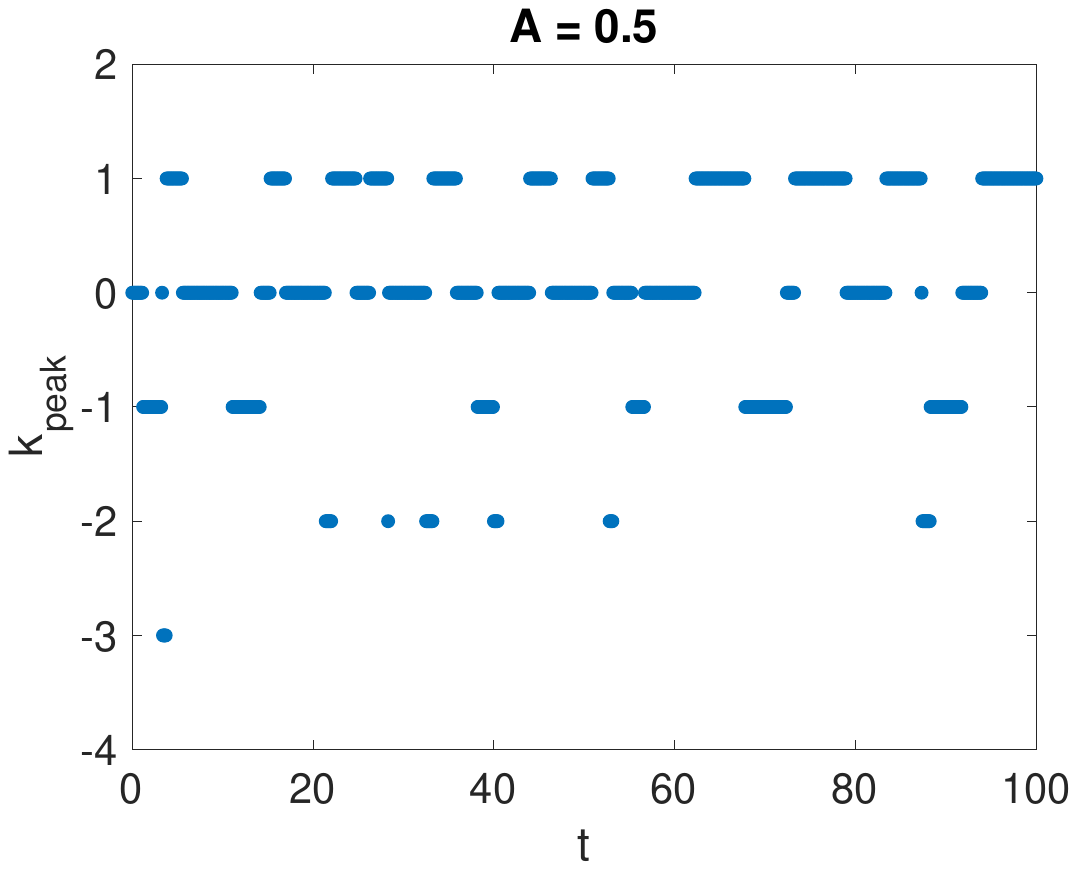}
\includegraphics[width=.35\textwidth]{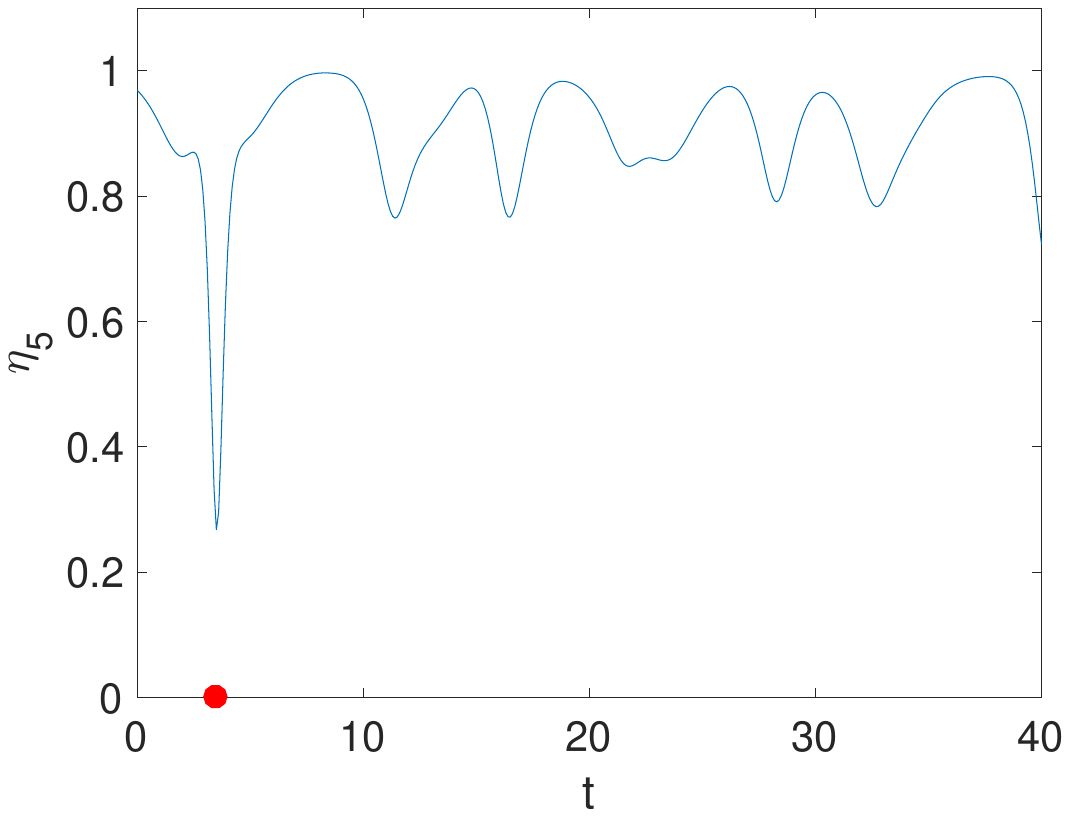}
  }
    \vspace{-48pt}
  \centerline{(a)\hspace{0.33\textwidth}(b)\hspace{0.33\textwidth}(c)}
  \caption{V-HONLS dynamics with SPB initial data and $\Gamma = 0.002$: (a) surface evolution, (b) Fourier spectral peak $k_{\mathrm{peak}}$, and (c) five-mode carrier-centered Fourier-energy fraction $\eta_5$.  Red bullets indicate the
    strong focusing events described in Section~5.2. The recovery of $\eta_5$ to large values between focusing events shows that the viscous evolution remains strongly concentrated in the central Fourier modes despite the independently established Floquet spectral reconfigurations.}
\label{fig:visc9}
\end{figure}

Following the common early-time evolution, the viscous interaction phases no longer settle into the comparatively persistent large-scale pattern observed in the NLD-HONLS case. As shown in Figure~\rf{fig:visc10}, both  $\psi_1$ and $\psi_2$ undergo repeated changes in drift tendency over the later evolution.
In contrast to NLD-HONLS, the two phases develop intervals of opposing drift rather than retaining a common long-time direction of evolution.

The relation between the phase changes and strong focusing is also less systematic. Although some local variations occur near the marked focusing events, the larger changes in the phase trajectories are not consistently tied to those events, unlike the focusing-associated behavior observed in the two-mode benchmark and the NLD-HONLS evolution. Thus, the viscous phase dynamics repeatedly change character rather than remaining organized around the recurrent focusing cycles.
\begin{figure}[htp!]
  \centerline{
  \includegraphics[width=.35\textwidth]{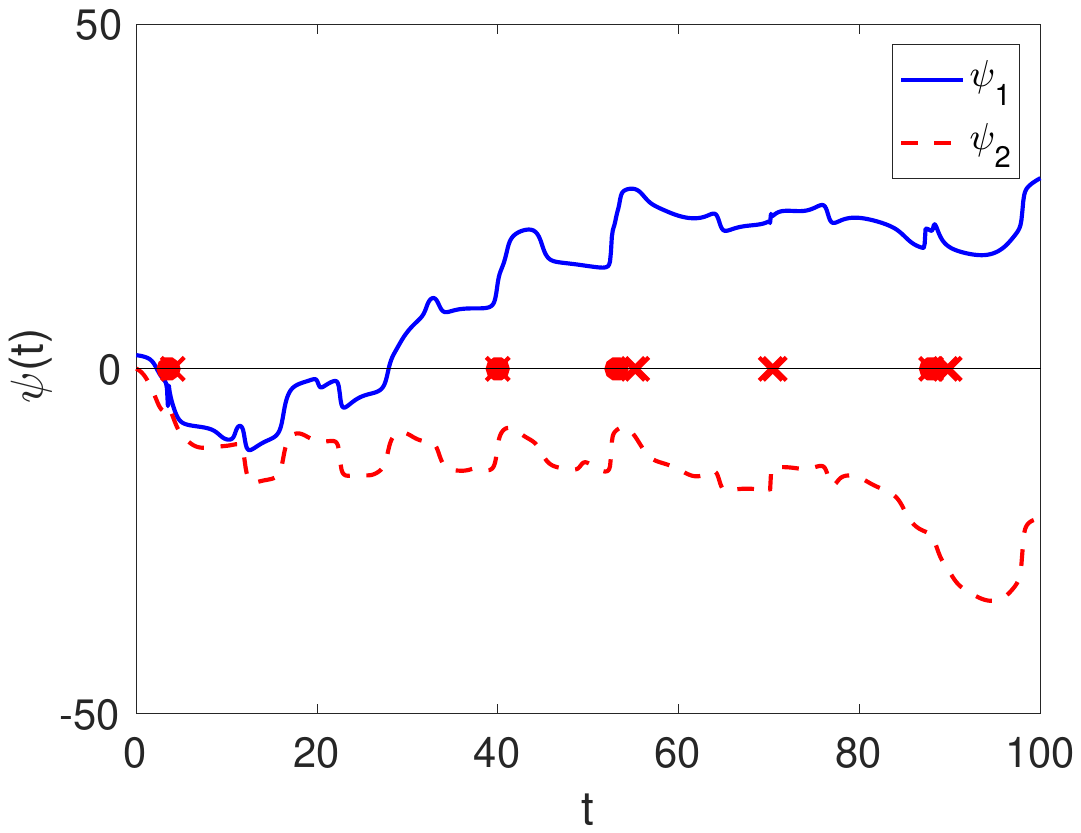}
  }
    \vspace{-48pt}
   \caption{V-HONLS principal   interaction phases $\psi_1$ and $\psi_2$
    for $\Gamma = 0.002$.
    Red bullets indicate strong focusing events defined in Section 5.2,
    while red ``$\times$'' symbols mark Floquet spectral-reconfiguration events.
    The two phases exhibit repeated changes in drift tendency, including intervals of opposing drift, and the larger local changes are not
    consistently associated with the marked focusing or Floquet spectral reconfiguration  events.}
\label{fig:visc10}
\end{figure}

These interaction-phase dynamics occur within the repeatedly reconfiguring Floquet regime established in Section~2. The viscous spectrum undergoes repeated critical-point crossings and band reconnections, with the spectra near
$t \approx 54.7$ in Figure~\rf{VHONLS_spec} providing one representative example. Thus, the repeated changes in phase-drift tendency occur while the Floquet-band configuration itself is also repeatedly reconfigured.

Comparison with the finite-gap benchmarks emphasizes that cumulative phase drift alone does not distinguish the two dissipative regimes. Under nonlinear mean-flow damping, the interaction phases retain a comparatively persistent large-scale pattern whose larger changes remain tied to recurrent focusing while the Floquet-band configuration persists. Under viscous damping, the phase trends change repeatedly, including intervals of opposing drift, without a comparable systematic association with recurrent focusing, within a regime of repeated Floquet-band reconfiguration.
The Fourier-energy diagnostics further show that this contrast is not explained simply by broader Fourier-mode excitation in the viscous system, since the central five modes continue to contain a large fraction of the total Fourier energy even as the nonlinear Floquet spectrum reconfigures.

This contrast is consistent with the mechanism  identified in
Section~4.3. Under nonlinear mean-flow damping, the  term $-\kappa_j \sin(\psi_j)$ provides direct restoring feedback on $psi_j$, opposing deviations  from the neighboring phase-aligned state while still allowing cumulative phase evolution under the remaining interaction terms.
This restoring feedback offers a possible explanation for the comparatively persistent NLD-HONLS phase dynamics, in which the two phases retain a common overall drift tendency and their larger local changes remain associated with recurrent strong focusing.
 In the viscous system, no corresponding phase-dependent restoring term exists; damping instead alters the modal amplitudes, which indirectly reweights the terms governing $\psi_1$ and $\psi_2$,
consistent with their repeatedly changing and at times opposing drift tendencies.

\section{Conclusion}

In this work, we investigated the interaction dynamics underlying the contrasting nonlinear Floquet spectral behavior previously observed in viscous and nonlinear mean-flow damped higher-order nonlinear Schr\"odinger systems. A reduced
five-mode carrier-sideband system was derived to identify how the two dissipative mechanisms enter the dominant modal interactions. In the viscous system, dissipation acts through modewise decay and does not contribute explicitly to the principal interaction-phase equations. Nonlinear mean-flow damping instead acts through the coupled mean-flow interaction terms and, in the carrier-sideband regime, generates direct phase-dependent contributions of the form
\be
-\kappa_j\sin(\psi_j).
\ee
This term provides nonlinear  restoring feedback on $\psi_j$, whereas no corresponding phase-dependent restoring contribution is present under viscous damping.
The five-mode system is therefore used here as a mechanism-identification model rather than as a quantitatively complete representation of the full PDE or its nonlinear Floquet spectrum. The Fourier-energy assessment further shows that the five retained modes remain energetically important over the carrier--sideband regime, while strong focusing temporarily excites additional nearby harmonics.

The NLS benchmark solutions establish an important point for interpreting these interaction phases. Substantial cumulative phase drift and markedly unequal evolution of $\psi_1$ and $\psi_2$
can coexist with recurrent modulation and an unchanged finite-gap Floquet configuration.
The two-mode benchmark further shows that pronounced local phase changes may remain tied to recurrent strong focusing events without indicating spectral reconfiguration.
Bounded or nearly stationary interaction phases are therefore not required for recurrent organization. Their evolution must be interpreted together with the modulation dynamics and nonlinear spectral configuration.

The full-PDE diagnostics reveal distinct phase behavior after the common early-time evolution of the two dissipative systems. Under nonlinear mean-flow damping, one interaction phase remains comparatively confined while the other undergoes substantially greater cumulative drift, but the two interaction phases $\psi_j$ retain predominantly the same long-time direction of evolution.
The more pronounced local phase changes remain associated with recurrent
strong focusing events. This occurs while the Floquet bands expand and drift without critical-point crossings or band reconfiguration. Under viscous damping, by contrast, both interaction phases undergo repeated changes in their large-scale drift tendencies, including intervals in which
$\psi_1$ and $\psi_2$ evolve in opposing directions.
Their larger phase  changes are not systematically tied to recurrent focusing and occur within the previously identified regime of repeated critical-point crossings and Floquet-band reconfiguration. The viscous solution nevertheless repeatedly returns to a strongly five-mode concentrated Fourier state, showing that Floquet reconfiguration is not simply a consequence of broad Fourier-mode excitation.
 
 Taken together, the reduced and full-PDE descriptions provide complementary but distinct information. The Fourier-energy diagnostics identify the modes carrying the dominant energy, while the reduced equations reveal how dissipation acts on the nonlinear interactions among those modes.
 In particular, the contrasting Floquet evolutions cannot be attributed simply to differences in Fourier-mode concentration.
 Instead, the reduced system identifies a structural distinction between the two damping mechanisms.
Nonlinear mean-flow damping acts directly through the coupled carrier-sideband mean-flow interactions and provides a restoring feedback on the principal interaction phases $\psi_j$, whereas viscous damping acts through modewise amplitude decay.
 
 The full-PDE phase dynamics are consistent with this distinction:  NLD-HONLS retains a comparatively persistent, focusing-associated phase evolution, whereas the viscous phase dynamics repeatedly change character and can develop opposing drift. This agreement is interpreted at the mechanism level and does not imply that the reduced model alone determines the nonlinear spectral evolution.

 \renewcommand{\thesection}{\Alph{section}}
\setcounter{section}{0}

\section{Supporting Fourier and interaction-phase calculations}
  \subsection{Fourier representation of the reduced interaction terms}

Substituting the five-mode truncation
\be
u(x,t)=\sum_{n=-2}^{2}A_n(t)e^{in\mu x},
\qquad
\mu=\frac{2\pi}{L},
\ee
into the governing equation \rf{DHONLS}, we derive the reduced Fourier interaction coefficients appearing in Section~3.
Since
\be
u_x=\sum_{n=-2}^{2}in\mu A_n e^{in\mu x},
\qquad
u_{xxx}=\sum_{n=-2}^{2}-in^3\mu^3 A_n e^{in\mu x},
\ee
the linear dispersive and viscous contributions act diagonally on the Fourier modes. Collecting coefficients of
\[
e^{in\mu x}
\]
gives the linear contribution
\be
\Omega_nA_n
=
\left(
\mu^2n^2-\frac{\epsilon}{2}\mu^3n^3
\right)A_n,
\ee
together with the viscous damping terms
\be
-i\Gamma A_n-2i\epsilon\Gamma\mu nA_n.
\ee

To derive the nonlinear interaction structure, first compute
\be
|u(x,t)|^2
=
u(x,t)\overline{u(x,t)}
=
\sum_{p=-4}^{4}Q_p(t)e^{ip\mu x},
\ee
where
\be
Q_p
=
\sum_{\substack{k,\ell=-2\\k-\ell=p}}^{2}
A_k\overline{A_\ell},
\qquad
p=-4,\dots,4.
\ee
Since \(|u|^2\) is real-valued,
\be
Q_{-p}=\overline{Q_p}.
\ee
The local cubic term becomes
\be
|u|^2u
=
\sum_{n=-2}^{2}C_n e^{in\mu x},
\ee
with interaction coefficients
\be
C_n
=
\sum_{\substack{k,\ell=-2\\n-k+\ell\in[-2,2]}}^{2}
A_k\overline{A_\ell}A_{n-k+\ell},
\qquad
n=-2,\dots,2.
\ee
These coefficients define the standard carrier-sideband four-wave interaction structure inherited from the NLS dynamics.

Next consider the nonlocal mean-flow contribution. Differentiating the quadratic expansion gives
\be
(|u|^2)_x
=
\sum_{p=-4}^{4}ip\mu Q_p e^{ip\mu x}.
\ee
Adopting the convention 
\be
\mathcal H(e^{ip\mu x})
=
+i\,\mathrm{sgn}(p)e^{ip\mu x},
\ee
the Hilbert-transform contribution becomes
\be
\mathcal H[(|u|^2)_x]
=
- \mu\sum_{p=-4}^{4}|p|Q_p e^{ip\mu x}.
\ee
Multiplying by $u$, expanding in Fourier modes,  and retaining only the modes $n = -2, ...2$  yields
\be
u\mathcal H[(|u|^2)_x]
=
- \mu\sum_{n=-2}^{2}M_n e^{in\mu x},
\ee
where
\be
M_n
=
\sum_{\substack{p=-4\\n-p\in[-2,2]}}^{4}
|p|Q_pA_{n-p},
\qquad
n=-2,\dots,2.
\ee
Finally, the self-steepening term gives
\be
|u|^2u_x
= i\mu
\sum_{n=-2}^{2}D_n e^{in\mu x},
\ee
with
\be
D_n
=
\sum_{\substack{p=-4\\n-p\in[-2,2]}}^{4}
(n-p)Q_pA_{n-p},
\qquad
n=-2,\dots,2.
\ee

Together, the coefficients \(C_n\), \(M_n\), and \(D_n\) define the nonlinear interaction structure of the reduced five-mode system. The cubic term generates the fundamental carrier-sideband four-wave exchange, the mean-flow term reweights the same interaction structure through nonlocal modulation, and the self-steepening term introduces mode-dependent asymmetry into the interaction dynamics.

\subsection{Amplitude-phase equations}

Writing the retained Fourier modes in amplitude-phase form,
\be
A_n(t)=r_n(t)e^{i\phi_n(t)},
\qquad
r_n\ge0,
\ee
substituting into the  five-mode equations derived in Section~3,
and separating real and imaginary parts yields coupled amplitude and phase equations.

For each retained mode \(n=-2,\dots,2\),
\be
\bdot A_n
=
(\bdot r_n+i r_n\bdot\phi_n)e^{i\phi_n}.
\ee
Substituting the reduced Fourier equations and multiplying by \(e^{-i\phi_n}\) gives
\begin{align}
\bdot r_n+i r_n\bdot\phi_n
={}&
-i\Omega_n r_n
+2i e^{-i\phi_n}C_n
-2i\epsilon\mu e^{-i\phi_n}M_n
\nonumber\\
&\quad
+2\beta\epsilon\mu e^{-i\phi_n}M_n
+8i\epsilon\mu e^{-i\phi_n}D_n
-\Gamma r_n
-2\epsilon\Gamma\mu n r_n.
\end{align}
Separating real and imaginary parts yields the amplitude equations
\begin{align}
\bdot r_n
={}&
-2\Im\!\left(e^{-i\phi_n}C_n\right)
+2\epsilon\mu
\Im\!\left(e^{-i\phi_n}M_n\right)
\nonumber\\
&\quad
+2\beta\epsilon\mu
\Re\!\left(e^{-i\phi_n}M_n\right)
-8\epsilon\mu
\Im\!\left(e^{-i\phi_n}D_n\right)
\nonumber\\
&\quad
-\Gamma r_n
-2\epsilon\Gamma\mu n r_n,
\label{appendix_amp_eq}
\end{align}
and the phase equations
\begin{align}
r_n\bdot\phi_n
={}&
-\Omega_n r_n
+2\Re\!\left(e^{-i\phi_n}C_n\right)
-2\epsilon\mu
\Re\!\left(e^{-i\phi_n}M_n\right)
\nonumber\\
&\quad
+2\beta\epsilon\mu
\Im\!\left(e^{-i\phi_n}M_n\right)
+8\epsilon\mu
\Re\!\left(e^{-i\phi_n}D_n\right).
\label{appendix_phase_eq}
\end{align}
Equivalently, dividing the phase equations by \(r_n\neq0\) gives
\begin{align}
\bdot\phi_n
={}&
-\Omega_n
+\frac{2}{r_n}\Re\!\left(e^{-i\phi_n}C_n\right)
-\frac{2\epsilon\mu}{r_n}
\Re\!\left(e^{-i\phi_n}M_n\right)
\nonumber\\
&\quad
+\frac{2\beta\epsilon\mu}{r_n}
\Im\!\left(e^{-i\phi_n}M_n\right)
+\frac{8\epsilon\mu}{r_n}
\Re\!\left(e^{-i\phi_n}D_n\right).
\label{appendix_phase_eq_divided}
\end{align}

Thus the dissipative contributions enter the amplitude and phase dynamics differently. The viscous terms contribute directly through diagonal modewise amplitude decay, while the nonlinear mean-flow damping contributes to both the amplitude and phase dynamics through the interaction terms involving \(M_n\).

\subsection{Exact interaction-phase equations}

We now derive the exact evolution equations for the principal interaction phases introduced in Section~4. The dominant carrier-sideband interaction phases are defined by
\be
\psi_j=\phi_j+\phi_{-j}-2\phi_0,
\qquad
j=1,2.
\ee
Differentiating gives
\be
\bdot\psi_j
=
\bdot\phi_j+\bdot\phi_{-j}-2\bdot\phi_0.
\ee
Substituting the phase equations derived in Appendix~A.2 yields
\be
\bdot\psi_j
=
\Delta_j
+
\mathcal{C}_j
+
\mathcal{M}_j
+
\mathcal{D}_j,
\qquad
j=1,2.
\ee
The linear phase-mismatch contributions are
\be
\Delta_j
=
-\Omega_j-\Omega_{-j}+2\Omega_0.
\ee
The  cubic interaction contributions are
\be
\mathcal{C}_j
=
\frac{2}{r_j}
\Re\!\left(e^{-i\phi_j}C_j\right)
+
\frac{2}{r_{-j}}
\Re\!\left(e^{-i\phi_{-j}}C_{-j}\right)
-
\frac{4}{r_0}
\Re\!\left(e^{-i\phi_0}C_0\right).
\ee
The mean-flow contributions are
\begin{align}
\mathcal{M}_j
={}&
-\frac{2\epsilon\mu}{r_j}
\Re\!\left(e^{-i\phi_j}M_j\right)
-\frac{2\epsilon\mu}{r_{-j}}
\Re\!\left(e^{-i\phi_{-j}}M_{-j}\right)
\nonumber\\
&\quad
+
\frac{4\epsilon\mu}{r_0}
\Re\!\left(e^{-i\phi_0}M_0\right)
\nonumber\\
&\quad
+
\frac{2\beta\epsilon\mu}{r_j}
\Im\!\left(e^{-i\phi_j}M_j\right)
+
\frac{2\beta\epsilon\mu}{r_{-j}}
\Im\!\left(e^{-i\phi_{-j}}M_{-j}\right)
\nonumber\\
&\quad
-
\frac{4\beta\epsilon\mu}{r_0}
\Im\!\left(e^{-i\phi_0}M_0\right).
\end{align}
Finally, the self-steepening contributions are
\be
\mathcal{D}_j
=
\frac{8\epsilon\mu}{r_j}
\Re\!\left(e^{-i\phi_j}D_j\right)
+
\frac{8\epsilon\mu}{r_{-j}}
\Re\!\left(e^{-i\phi_{-j}}D_{-j}\right)
-
\frac{16\epsilon\mu}{r_0}
\Re\!\left(e^{-i\phi_0}D_0\right).
\ee

These equations retain the full decomposition  into linear phase mismatch, cubic four-wave exchange, mean-flow interaction, dissipative mean-flow corrections, and self-steepening contributions.

\subsection{Carrier-sideband asymptotics}

We now derive the leading carrier-sideband asymptotics associated with the nonlinear mean-flow damping contribution appearing in the NLD-HONLS interaction-phase equations of Section~4.
 The distinguished dissipative contribution arises from the \(\beta\)-dependent mean-flow terms appearing in the exact interaction-phase equations,
\be
\frac{2\beta\epsilon\mu}{r_j}
\Im\!\left(e^{-i\phi_j}M_j\right)
+
\frac{2\beta\epsilon\mu}{r_{-j}}
\Im\!\left(e^{-i\phi_{-j}}M_{-j}\right)
-
\frac{4\beta\epsilon\mu}{r_0}
\Im\!\left(e^{-i\phi_0}M_0\right),
\qquad
j=1,2.
\ee
Under the carrier-sideband hierarchy
\be
r_0 \gg r_{\pm1}\gg r_{\pm2},
\ee
the dominant contributions to \(M_n\) arise from the principal carrier-sideband interaction products.

For
\be
\psi_j=\phi_j+\phi_{-j}-2\phi_0,
\qquad
j=1,2,
\ee
the dominant quadratic correlations are
\be
Q_j\sim A_0\overline{A_{-j}},
\qquad
Q_{-j}\sim A_j\overline{A_0}.
\ee
From the definition
\be
M_n
=
\sum_{p=-4}^{4}|p|Q_pA_{n-p},
\ee
the dominant contribution to \(M_j\) arises from \(p=j\), giving
\be
M_j\sim jQ_jA_0.
\ee
Keeping only the leading phase-dependent terms contributing to
\[
\Im\!\left(e^{-i\phi_n}M_n\right),
\]
gives
\[
Q_j \sim r_0r_{-j}e^{i(\phi_0-\phi_{-j})},
\]
and therefore
\[
M_j \sim jr_0^2r_{-j}e^{i(2\phi_0-\phi_{-j})}.
\]
Consequently,
\be
e^{-i\phi_j}M_j
\sim
jr_0^2r_{-j}e^{-i\psi_j},
\ee
so that
\be
\Im\!\left(e^{-i\phi_j}M_j\right)
\sim
-jr_0^2r_{-j}\sin(\psi_j).
\ee
Similarly,
\be
\Im\!\left(e^{-i\phi_{-j}}M_{-j}\right)
\sim
-jr_0^2r_j\sin(\psi_j).
\ee
For the carrier mode, the dominant contributions arise from \(p=\pm j\),
\be
M_0
\sim
jQ_jA_{-j}
+
jQ_{-j}A_j,
\ee
which gives
\be
\Im\!\left(e^{-i\phi_0}M_0\right)
\sim
2jr_0r_jr_{-j}\sin(\psi_j).
\ee
Substituting these leading contributions into the \(\beta\)-dependent part of the interaction-phase equation yields
\be
-\kappa_j\sin(\psi_j),
\ee
where
\be
\kappa_j
=
2\beta\epsilon\mu
\left[
jr_0^2
\left(
\frac{r_{-j}}{r_j}
+
\frac{r_j}{r_{-j}}
\right)
+
4jr_jr_{-j}
\right],
\qquad
j=1,2.
\ee

Equivalently,
\be
\kappa_1
=
2\beta\epsilon\mu
\left[
r_0^2
\left(
\frac{r_{-1}}{r_1}
+
\frac{r_1}{r_{-1}}
\right)
+
4r_1r_{-1}
\right],
\ee
while
\be
\kappa_2
=
2\beta\epsilon\mu
\left[
2r_0^2
\left(
\frac{r_{-2}}{r_2}
+
\frac{r_2}{r_{-2}}
\right)
+
8r_2r_{-2}
\right].
\ee

These asymptotic reductions isolate the leading interaction-phase contributions generated by nonlinear mean-flow damping within the dominant interaction regime.

\section{Reference NLS solutions and benchmark initial data}
\subsection{One-mode and two-mode finite-gap benchmark solutions}
The following benchmark solutions are used throughout Section~5.3 to study the interaction-phase dynamics and to verify the numerical accuracy of the HONLS integrator. Their observable dynamics are effectively lower-dimensional due to their underlying spatial and spectral symmetries. In particular, the solutions possess even spatial symmetry, and their associated Floquet spectra are symmetric with respect to both the real and imaginary axes.

The ``one-mode'' finite-gap solution  is initialized using the exact three-phase finite-gap solution given by \cite{AKE}:
 \be 
  u_0(x,t) = \frac{\kappa}{\sqrt{2}} e^{\ri t}\frac{\cn\left(\sqrt{\frac{1+\kappa}{2}}x, k\right)\,
    \cn(t,\kappa)+
    \ri\sqrt{1+\kappa}\,\dn\left(\sqrt{\frac{1+\kappa}{2}}x,k\right)\,\sn(t,\kappa)}
  {\sqrt{1+\kappa}\,\dn\left(\sqrt{\frac{1+\kappa}{2}}x,k\right) -     \cn\left(\sqrt{\frac{1+\kappa}{2}}x,k\right)\,
    \dn(t,\kappa)},
  \label{3phase}
  \ee
where $0 < \kappa <1$,
and $k = \sqrt{\frac{1-\kappa}{1+\kappa}}$.
  The spatial period, $L_x $, and the temporal period of the modulated phase, $L_t $,
    are functions of the complete elliptic integrals of
    the first kind,   ${{\cal K}_x(k)}$ and ${{\cal K}_t(\kappa)}$,
    respectively.
    For generic $0 < \kappa <1$, the corresponding Floquet spectral curve has three finite spectral bands and genus $g = 2$. Thus, this is a genus-two,
    three-phase solution, where the phase of the background Stokes wave is
    included in the three-phase count.
Its observable dynamics are dominated by a single recurrent modulation mode, as illustrated in Figure \rf{fig:1UM}. Throughout the paper, this solution is referred to as the one-mode finite-gap benchmark solution.

The second benchmark solution, the ``two-mode'' finite gap solution, is initialized using the perturbed data,
\be
u(x,0) = a(1 + \delta \cos(\mu x) + \delta \cos(2 \mu x)),
\label{5phase}
\ee
with parameters chosen so that the resulting evolution organizes into a symmetric recurrent two-mode modulation structure that  closely approximates a five-phase finite-gap solution, as observed in Figure \rf{fig:2UM}. A five-phase solution corresponds to a genus-four spectral curve, with four nontrivial finite-gap phases together with the phase of the background Stokes wave.
Although the initial condition \rf{5phase} does not generate an exact finite-gap solution, the additional spectral bands generated by the perturbation remain
spectrally small  over the time intervals considered here, so that the evolution is well approximated by the dominant five-phase finite-gap state.
  \subsection{Spatially periodic breather solution}

 The spatially periodic breathers are homoclinic orbits of unstable Stokes waves $u_a(t)$ and can be explicitly constructed by means of B\"acklund transformations \cite{SZ1987}.
 In the case of a Stokes wave with two or more unstable
modes, the B\"acklund formula can be iterated algebraically to provide
a complete representation of the homoclinic manifold of $u_a(t)$. 

The two-mode SPB over an unstable Stokes wave is given by: 
\be
U(x,t;\rho,
\tau)=a e^{2\ri a^2 t} \frac{N(x,t;\rho, \tau)}{D(x,t;\rho, \tau)},
\label{comboSPB}
\ee
where

\[
\begin{array}{rl}
& N(x,t;\rho, \tau)  = 4 (\sin^2 p \cos 2p +\sin^2 q \cos 2q)  \\ & + 4\sin
  p(\sin^2 q -\sin^2 p) \cos 2q \cos\left({2k_1}x+\beta
  \right) \mbox{sech}(\rho -\sigma t) \\ & -2  \sin 2p \sin 2q \, \tanh (\tau -\delta t) \tanh
  (\rho -\sigma t) \\ &-4\sin q(\sin^2 q -\sin^2
  p)\cos\left({2k_2}x+\gamma \right) \mbox{sech}(\tau -\delta
  t)  \\
 & -4\sin p \sin q( \cos^2 p +  \cos^2 q) 
\cos \left(2k_1x+\beta \right) \cos \left(2k_2x+\gamma \right) 
 \mbox{sech}(\rho -\sigma t) \mbox{sech}(\tau -\delta t) \\\
&  - 4 \ri \sin 2p (\sin^2 q -\sin^2 p) \tanh (\rho -\sigma t) 
 4 \ri \sin 2q (\sin^2 q -\sin^2 p) \tanh (\tau -\delta t)  \\
&  + 4 \ri \sin p \sin 2q (\sin^2 q -\sin^2 p) \tanh (\tau -\delta t) \mbox{sech}(\rho -\sigma t)
\cos\left(2k_1x+\beta \right)  \\
&-  4 \ri \sin 2p \sin q (\sin^2 q -\sin^2 p) \tanh (\rho -\sigma t) \mbox{sech}(\tau -\delta t)
\cos\left(2k_2x+\gamma \right)  \\
& - 2 \sin 2p \sin 2q 
\sin \left(2k_1 x+\beta \right) \sin \left(2k_2 x+\gamma \right) 
 \mbox{sech}(\rho -\sigma t) \mbox{sech}(\tau -\delta t).
\end{array}
\]

and
\[
\begin{array}{rl}
 & D(x,t;\rho, \tau)  =  4(\sin^2p\cos^2q+\sin^2q\cos^2p) \\ 
 & + 4\sin q(\sin^2 q
  -\sin^2 p)\cos\left(2k_2x+\gamma \right) \mathrm{sech}(\tau
  -\delta t)\\  
 &-4\sin p(\sin^2 q - \sin^2 p) \cos\left(2k_1x+\beta \right) \mbox{sech}(\rho -\sigma t) \\
&- 4\sin p \sin q  (\cos^2 p + \cos^2 q)
\cos\left(2k_1x+\beta \right)
\cos\left(2k_2x+\gamma \right) \mbox{sech}(\rho -\sigma t)
\mathrm{sech}(\tau -\delta t)  \\
& -2 \sin 2p \sin 2q \, \tanh (\tau -\delta t) \tanh (\rho -\sigma t) \\ 
& -2 \sin 2p \sin 2q 
\sin \left(2k_1x+\beta \right) \sin \left(2k_2x+\gamma \right) 
 \mbox{sech}(\rho -\sigma t) \mbox{sech}(\tau -\delta t). 
\end{array}
\]
The above formula  contains two spatial
modes with wavenumbers $k_1=\pi/L$ and $k_2=2\pi/L$, and depends on two real parameters, $\rho$ and $\tau$, which determine the time at which the first and second modes are excited, respectively.
Figure~\rf{Coal_U2}(a)
shows 
the two-mode SPB with maximal amplitude, corresponding to the parameter choice
$\rho = 0$ and $\tau = 0$, where   both modes are simultaneously excited. In this case the solution exhibits a single localized focusing event and 
then asymptotically decays to the Stokes wave as $t\to\pm\infty$.
  \begin{figure}[htp]
  \centerline{
 \includegraphics[width=0.3\textwidth]{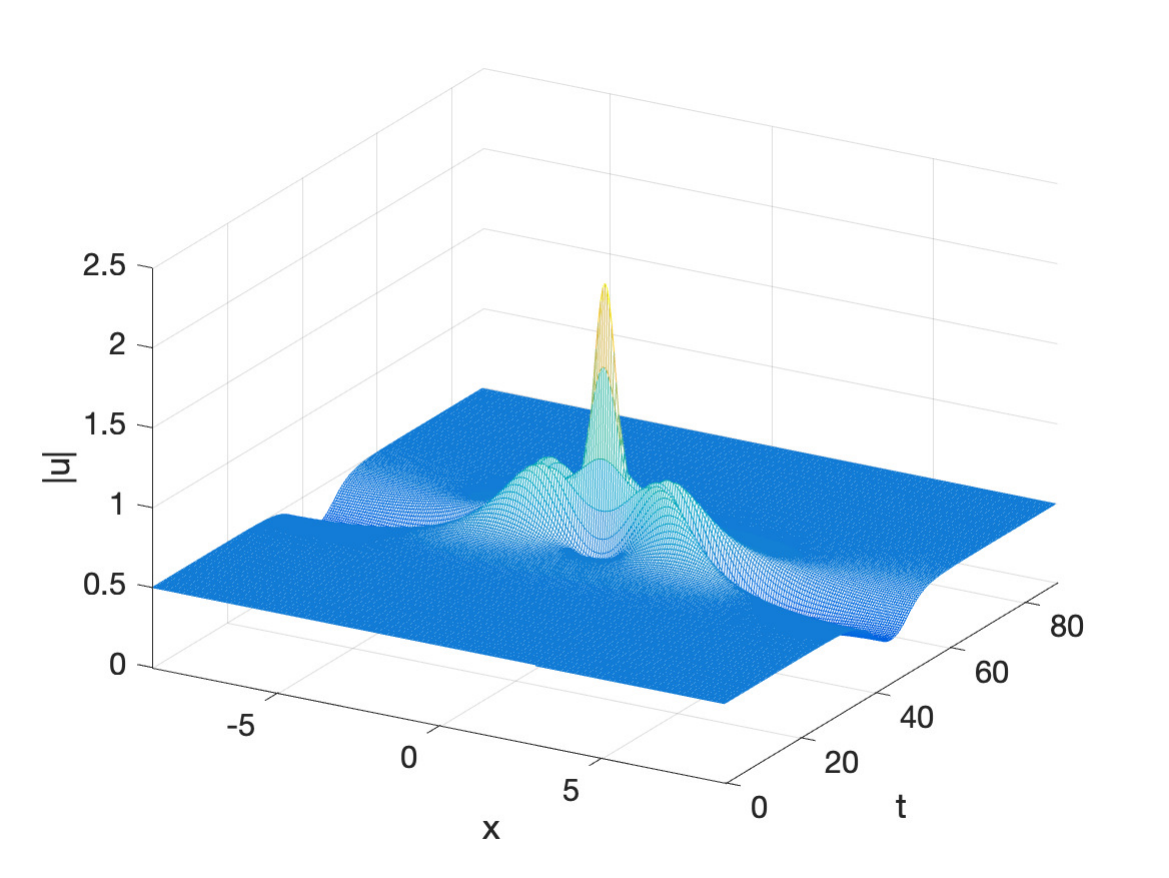}
   \includegraphics[width=0.3\textwidth]{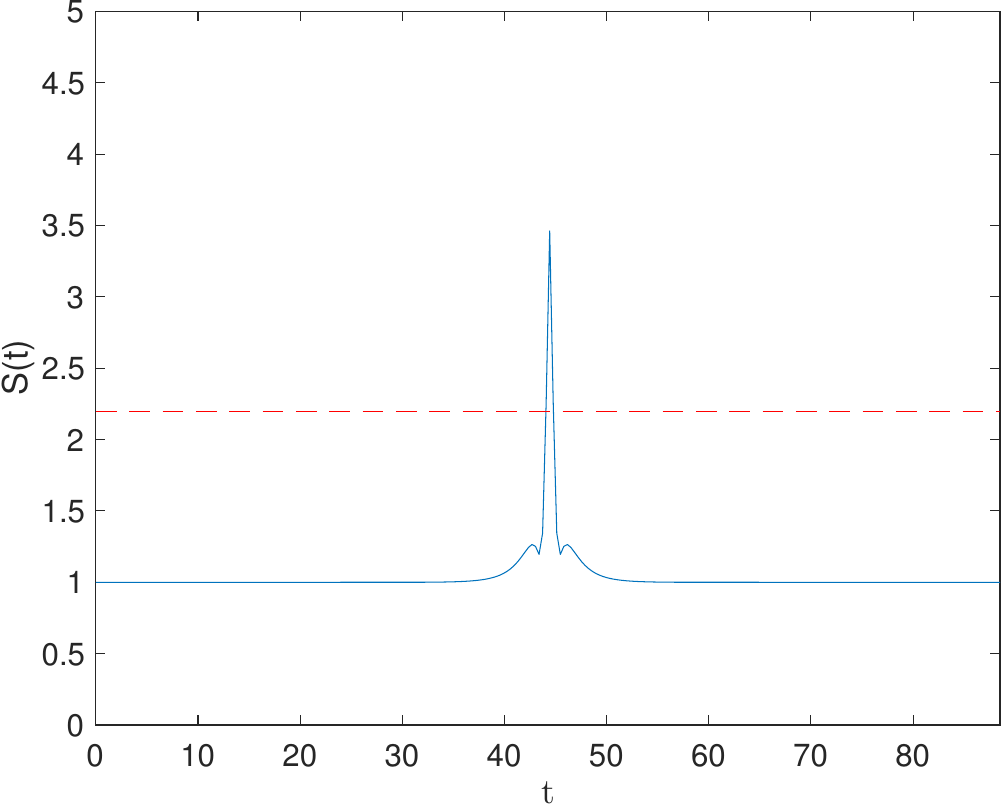}
  }
  \centerline{(a)\hspace{0.33\textwidth}(b)}
  \caption{Coalesced two-mode SPB for $\rho = 0$ and $\tau =0$: (a)  $|U(x,t)|$ and (b) S(t) for $0\le t \le 80$.}
  \label{Coal_U2}
  \end{figure}
  The corresponding strength plot in Figure~\rf{Coal_U2}(b) confirms the
  occurrence  of a single
  rogue wave event, consistent with this decay behavior.
The Floquet spectrum for the  two-mode SPB  
is shown in
 Figure~\rf{early-spec}(a).

\section*{Acknowledgments}
I gratefully acknowledge support for this project by the Simons Foundation through award \#527565.

\section*{Additional Information}
\begin{itemize}
  \item The author does not have a conflict of interest to disclose.
\item Data sharing not applicable to this article as no datasets were generated or analysed during the current study.
\end{itemize}

\bibliographystyle{plain}

\begin{thebibliography}{10}
\bibitem{ablowitz_segur}
M.J. Ablowitz and H. Segur,
\newblock {\em Solitons and the inverse scattering transform}.
\newblock SIAM, Philadelphia, 1981.

\bibitem{ablowitz2011}
M.J. Ablowitz,
\newblock {\em Nonlinear dispersive waves: Asymptotic analysis and Solitons}.
\newblock Cambridge University Press, Cambridge, 2011.

\bibitem{AC1991}
M.J. Ablowitz and P.A. Clarkson.
\newblock {\em Solitons, Nonlinear Evolution Equations and Inverse Scattering
  Transform}.
\newblock Cambridge University Press, Cambridge, 1991.

\bibitem{AKNS}
M.~J. Ablowitz, D.~J. Kaup, A.~C. Newell, and H.~Segur.
\newblock The inverse scattering transform-fourier analysis for nonlinear
  problems.
\newblock {\em Studies in Applied Mathematics}, 53(4):249--315, 1974.

\bibitem{AHS1996}
M.J. Ablowitz, B.M. Herbst, and C.M. Schober.
\newblock Computational chaos in the nonlinear Schr\"odinger equation without homoclinic crossings.
\newblock{\em Physica A: Stat. Mech. Appl.}, 228: 212--235, 1996.

\bibitem{Benney1966}
D.~J. Benney.
\newblock Non-linear gravity wave interactions.
\newblock {\em Journal of Fluid Mechanics}, 14:577--584, 1962.

\bibitem{BenneyNewell1967}
D.~J. Benney and A.~C. Newell.
\newblock The propagation of nonlinear wave envelopes.
\newblock {\em Journal of Mathematics and Physics}, 46:133--139, 1967.

\bibitem{DubrovinNovikov}
B.~A. Dubrovin and S.~P. Novikov.
\newblock Periodic and conditionally periodic analogs of the many-soliton
  solutions of the korteweg-de vries equation.
\newblock {\em Soviet Physics JETP}, 40(6):1058--1063, 1975.

\bibitem{ForestGoedde}
M.~G. Forest and C.~G. Goedde.
\newblock Chaotic transport and integrable instabilities in a near-integrable, many-particle, Hamiltonian lattice.
\newblock {\em Physica D: Nonlinear Phenomena}, 53(4):1116--1138, 1993.

\bibitem{Hasselmann}
K.~Hasselmann.
\newblock On the nonlinear energy transfer in a gravity-wave spectrum. part 1.
  general theory.
\newblock {\em Journal of Fluid Mechanics}, 12(4):481--500, 1962.

\bibitem{ItsMatveev}
A.~R. Its and V.~B. Matveev.
\newblock Hill operators with finitely many lacunae.
\newblock {\em Functional Analysis and Its Applications}, 9(1):65--66, 1975.

\bibitem{Janssen}
P.~A. E.~M. Janssen.
\newblock {\em The Interaction of Ocean Waves and Wind}.
\newblock Cambridge University Press, Cambridge, 2004.

\bibitem{kodama}
Y.~Kodama.
\newblock {\em Solitons in Two-Dimensional Shallow Water}.
\newblock SIAM, Philadelphia, 2018.

\bibitem{McKeanVanMoerbeke}
H.~P. McKean and P.~van Moerbeke.
\newblock The spectrum of hill's equation.
\newblock {\em Inventiones Mathematicae}, 30(3):217--274, 1975.

\bibitem{OMB1986}
E.A. Overman~II, D.W. McLaughlin, and A.R. Bishop.
\newblock Coherence and chaos in the driven damped sine-gordon equation:
  measurement of the soliton spectrum.
\newblock {\em Physica D}, 19:1--41, 1986.

\bibitem{EFM1990}
 N. Ercolani, M.G. Forest, and D.W. McLaughlin.
\newblock Geometry of the modulational instability III: homoclinic orbits for the periodic Sine-Gordon equation.
\newblock {\em Physica D}, 43:349--384, 1990.

\bibitem{McLaughlinSchober}
D.~W. McLaughlin and C.~M. Schober.
\newblock Chaotic and homoclinic behavior for numerical discretizations of the
nonlinear schr\"odinger equation.
\newblock {\em Physica D}, 57(3):447--465, 1992.

\bibitem{Novikov}
S.~P. Novikov.
\newblock The periodic problem for the korteweg-de vries equation.
\newblock {\em Functional Analysis and Its Applications}, 8(3):236--246, 1974.

\bibitem{AS2002}
  M.J. Ablowitz and C.M. Schober.
\newblock Chaotic Dynamics in Nonlinear Waves: Computational and Physical.
\newblock {\em SIAM Lecture Series,  ``Collected Lectures on the Preservation of Stability under Discretization'', Eds. D. Esteep and S. Tavner}, PR109: 237--264, 2002.


\bibitem{Whitham}
G.~B. Whitham.
\newblock {\em Linear and Nonlinear Waves}.
\newblock Wiley, New York, 1974.

\bibitem{Zakharov}
V.E. Zakharov.
\newblock Stability of periodic waves of finite amplitude on the surface of a
  deep fluid.
\newblock {\em Journal of Applied Mechanics and Technical Physics},
  9(2):190--194, 1968.

\bibitem{CM2002}
S.M. Cox, P.C. Matthews.
\newblock Exponential time differencing for stiff systems,
\newblock {\em J. Comp. Phys.}, 176: 430--455, 2002.

\bibitem{AKE}
  N.N. Akhmediev, V.M. Eleonskii, N.E. Kulagin.
  \newblock Exact first-order solutions of the nonlinear Schr\"odinger equation.
  \newblock {\em Theor. Math. Phys. (USSR)}, 72: 809--818, 1987.

\bibitem{CG2016}
J.D. Carter and A.~Govan.
\newblock Frequency downshift in a viscous fluid.
\newblock {\em European Journal of Mechanics - B/Fluids}, 59:177--185, 2016.

\bibitem{FD2011}
F.~Fedele and D.S. Dutykh.
\newblock Hamiltonian form and solitary waves of the spatial dysthe equations.
\newblock {\em JETP Letters}, 94:840--844, 2011.

\bibitem{GT2011}
O.~Gramstad and K.~Trulsen.
\newblock Hamiltonian form of the modified nonlinear schr\"odinger equation for
  gravity waves on arbitrary depth.
\newblock {\em Journal of Fluid Mechanics}, 670:404--426, 2011.

\bibitem{IS2011}
A.~Islas and C.M. Schober.
\newblock Rogue waves and downshifting in the presence of damping.
\newblock {\em Nat. Hazards Earth Syst. Sci.}, 11:383--399, 2011.

\bibitem{SI2025}
C.M. Schober and A. Islas.
\newblock Soliton-like rogue wave dynamics in dissipative higher order
nonlinear Schr\"odinger Models: A Floquet spectral perspective.
\newblock {\em Studies in Applied Mathematics}, 155:10.1111/sapm.70150, 2025.

\bibitem{SI2022}
C.M. Schober and A. Islas.
\newblock Nonlinear damped spatially periodic breather and the emergence of
  soliton-like rogue waves.
\newblock {\em Physica D}, 438, 2022.

\bibitem{KO1995}
Y.~Kato and M.~Oikawa.
\newblock Wave number downshift in modulated wavetrain through a nonlinear
  damping effect.
\newblock {\em Journal of the Physical Society of Japan}, 64(12):4660--4669,
  1995.

\bibitem{SZ1987}
D.H. Sattinger and V.D. Zurkowski.
\newblock Gauge theory of bäcklund transformations.
\newblock {\em Physica D}, 26:225--250, 1987.

\bibitem{SI2021}
C.M. Schober and A.L. Islas.
\newblock On the stabilization of breather type solutions of the damped higher
  order nls.
\newblock {\em Front. Phys.}, 9, 2021.

\bibitem{UK1994}
Y.~Uchiyama and T.~Kawahara.
\newblock A possible mechanism for frequency downshift in nonlinear wave
  modulation.
\newblock {\em Wave Motion}, 20:99--110, 1994.

\bibitem{ZS1972}
V.E. Zakharov and A.B. Shabat.
\newblock Exact theory of two-dimensional self-focusing and one-dimensional
  self-modulation of waves in nonlinear media.
\newblock {\em Soviet Phys. JETP}, 34:62--69, 1972.

\bibitem{BF1967}
T.B. Benjamin and J.E. Feir.
\newblock The disintegration of wave trains on deep water. Part 1. Theory.
\newblock{\em J Fluid Mech}, 27:417--430, 1967.

\bibitem{OOSB2001}
M. Onorato, A. Osborne, M. Serio, and S. Bertone.
\newblock Freak waves in random oceanic sea states.
\newblock{\em Phys Rev Let}, 86:5831--5834, 2001.

\bibitem{EALBBK2020}
D. Eeltink, A. Armaroli, C. Luneau, H. Branger, M. Brunetti, and J. Kasparian.
\newblock Separatrix crossing and symmetry breaking in NLSE-like systems due to forcing and damping.
\newblock {\em Nonlinear Dynamics}, 102:2385--2398, 2020.

\bibitem{HW1995}
G. Haller and S. Wiggins.
\newblock Multi-pulse jumping orbits and homoclinic trees in a modal truncation of the damped-forced nonlinear Schr\"odinger equation.
\newblock {\em Physica D: Nonlinear Phenomena}, 85:311--347, 1995.

\bibitem{FCSR2025}
L. Fache, F. Copie, P. Surret, and S. Randoux.
\newblock Perturbed Nonlinear Evolution of Optical Soliton Gases: Growth and Decay in Integrable Turbulence.
\newblock {\em Phys. Rev. Lett.}, 135:157201, 2025.

\bibitem{FBD2024}
Loic Fache, Félicien Bonnefoy, Guillaume Ducrozet, {\em et al}.
\newblock Interaction of soliton gases in deep-water surface gravity waves.
\newblock {\em Phys. Rev. E}, 109:034207, 2024.

\bibitem{Trillo-etal2018}
A. Mussot, C. Naveau, M. Conforti, {\em et al}.
\newblock Fibre multiwave mixing combs reveal the broken symmetry of Fermi–Pasta–Ulam recurrence.
\newblock {\em Nature Photonics}, 12:303--308, 2018.
\end{thebibliography}

\end{document}